\documentclass[aps,prl,reprint,amsmath,amssymb,superscriptaddress,longbibliography]{revtex4-2}

\usepackage{graphicx}% Include figure files
\usepackage{dcolumn}% Align table columns on decimal point
\usepackage{bm}% bold math
\usepackage{amsmath}
\usepackage{amssymb}
\usepackage{xcolor}
\usepackage{xspace}
\usepackage{xfrac}
\usepackage{color,soul}
\usepackage[colorlinks=true,urlcolor=blue,citecolor=blue,linkcolor=blue,bookmarks=false,pdfstartview={FitH}]{hyperref}
\usepackage{subcaption}
\definecolor{darkgreen}{rgb}{0.0, 0.42, 0.10}

\DeclareMathOperator{\sgn}{sgn}

\begin{document}
\title{Current- and field-induced topology in twisted nodal superconductors}

\author{Pavel A. Volkov}
\email{pv184@physics.rutgers.edu}
\affiliation{Department of Physics and Astronomy, Center for Materials Theory, Rutgers University, Piscataway, NJ 08854, USA}
\affiliation{Department of Physics, Harvard University, Cambridge, Massachusetts 02138, USA}
\affiliation{Department of Physics, University of Connecticut, Storrs, Connecticut 06269, USA}
\author{Justin H. Wilson}
\affiliation{Department of Physics and Astronomy, and Center for Computation and Technology, Louisiana State University, Baton Rouge, LA 70803, USA}
\affiliation{Department of Physics and Astronomy, Center for Materials Theory, Rutgers University, Piscataway, NJ 08854, USA}
\author{Kevin Lucht}
\affiliation{Department of Physics and Astronomy, Center for Materials Theory, Rutgers University, Piscataway, NJ 08854, USA}
\author{J. H. Pixley}
\affiliation{Department of Physics and Astronomy, Center for Materials Theory, Rutgers University, Piscataway, NJ 08854, USA}
\affiliation{Center for Computational Quantum Physics, Flatiron Institute, 162 5th Avenue, New York, NY 10010} 
 \affiliation{Physics Department, Princeton University, Princeton, New Jersey 08544, USA}

\begin{abstract}
We show that interlayer current induces topological superconductivity in twisted bilayers of nodal superconductors. A bulk gap opens and achieves its maximum near a ``magic'' twist angle $\theta_\mathrm{MA}$. Chiral edge modes lead to a quantized thermal Hall effect at low temperatures. Furthermore, we show that an in-plane magnetic field creates a periodic lattice of topological domains with edge modes forming low-energy bands. We predict their signatures in scanning tunneling microscopy. Estimates for candidate materials indicate that twist angles $\theta\sim \theta_\mathrm{MA}$ are optimal for observing the predicted effects.
\end{abstract}

\maketitle

Controlling the Bogoliubov-de Gennes (BdG) excitations in superconductors (SC) is crucial for realizing many coveted quantum phases of matter.
For example, topologically nontrivial BdG bands~\cite{schnyder2008} hold the promise of hosting  exotic Majorana fermion excitations \cite{sato2017}
that can be used to perform topological quantum computation~\cite{sarma2015majorana}. 
However, despite many considered materials \cite{Nandkishore2012,feng2013,fischer2014,zhang2019multiple}
and nanostructure setups \cite{mourik2012signatures}, the controlled realization of topological phases of the BdG quasiparticles remains an open problem. 
Fundamentally, low-energy BdG quasiparticles are charge neutral combinations of particles and holes~\cite{kivelson1990,ronen2016}, making the electric-field based control used in various semiconductor applications ineffective.

\begin{figure}[h!]
	\includegraphics[width=\linewidth]{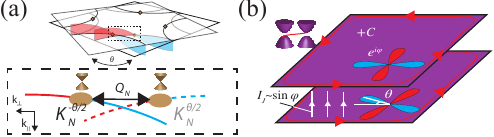}
	\caption{(a) Momentum-space schematic of a twisted nodal superconductor exemplified by a $d$-wave superconductor with a sign-changing gap (from blue to red). Near the nodes $(K_N$ and $\tilde K_N)$ the BdG quasiparticles of the two layers have a Dirac dispersion shifted by a vector $Q_N(=\theta K_N)$ with respect to one another. (b) Interlayer current leads to opening of a bulk $\mathbb{Z}$ topological gap with gapless chiral edge modes (Fig.~\ref{fig:gap}).}
	\label{fig:cart}
\end{figure}

Recently, a new paradigm in the engineering of correlated and topological phases has emerged, known as ``twistronics''~\cite{lopes2007,bistritzer2011,carr2017} or moir\'e materials~\cite{balents2020}, that utilizes stacking of two-dimensional materials with an interlayer rotation (i.e., twist as in Fig.~\ref{fig:cart}) to achieve novel properties. In particular, recent studies \cite{can2020hightemperature,tummuru2020chiral} have shown that twisted bilayers of nodal superconductors can spontaneously break time-reversal symmetry at certain twist angles (45$^\circ$ for d-wave superconductors) \cite{kuboki1996,sigrist1998}, potentially leading to topological states. The cuprates~\cite{can2020hightemperature} are such a candidate available in monolayer form~\cite{frank2019,yu2019}.
However, their topological properties are suppressed by the symmetry of the orbitals when twist angles are near $45^\circ$ \cite{song2022} (although incoherent tunneling have been suggested to reduce this effect \cite{haenel2022}).

\begin{figure}[h!]
	\includegraphics[width=\linewidth]{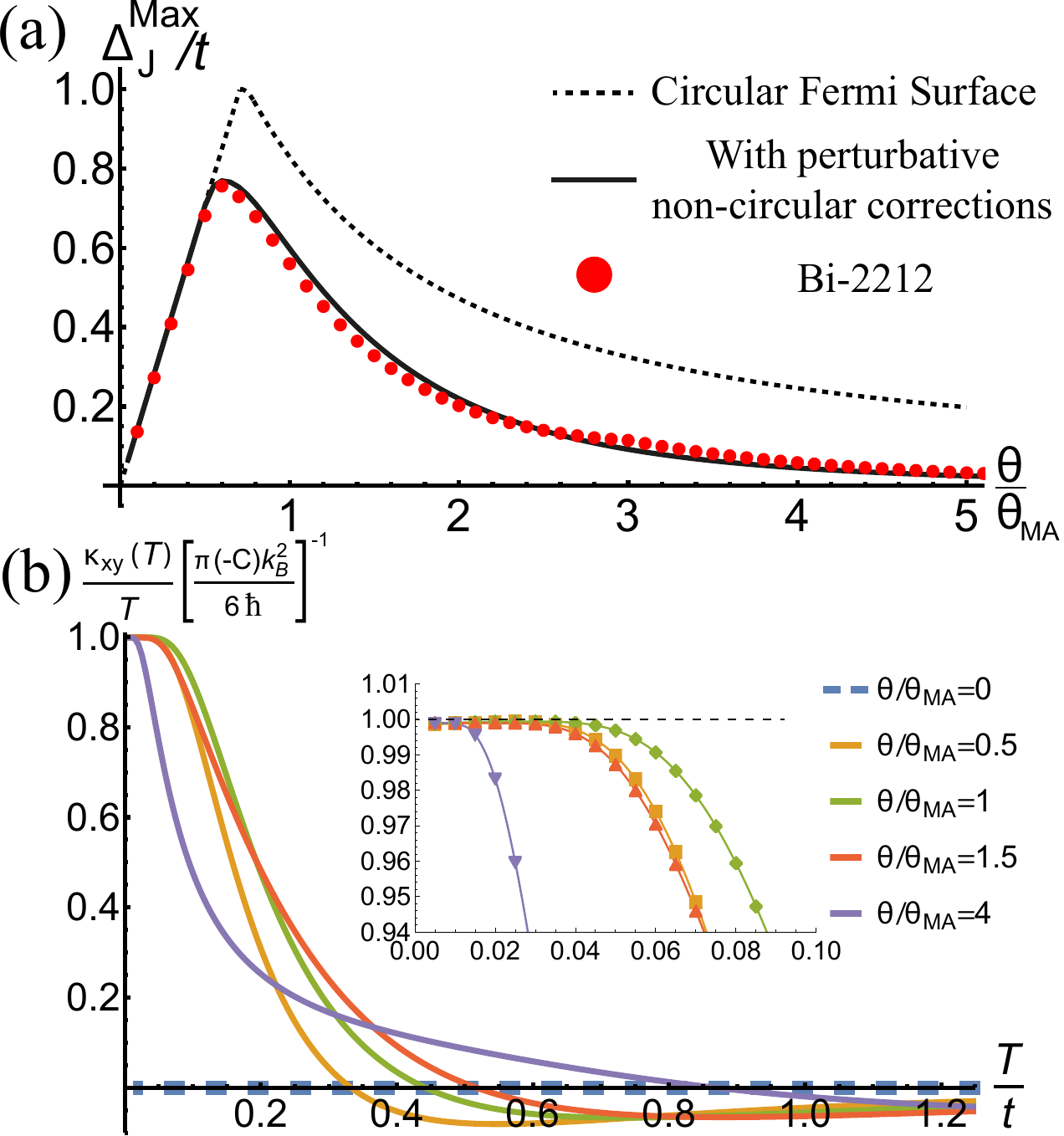}
	\caption{(a) Maximal value of the current-induced gap as a function of twist angle. Dashed line is for the circular Fermi surface (Eq.~\eqref{eq:curgap}), red dots - for a tight-binding model of Bi$_2$Sr$_2$CaCu$_2$O$_{8+ y}$ \cite{markiewicz2005,Note2}, green line - low-$\theta$ expansion for a non-circular Fermi surface \cite{Note2}. (b) Temperature dependence of the normalized thermal Hall conductivity for different twist angles and $\varphi=\pi/2$. For $\theta=0$, $\kappa_H(T)$ vanishes identically.}
	\label{fig:gap}
\end{figure}

\begin{figure}[h!]
	\includegraphics[width=\linewidth]{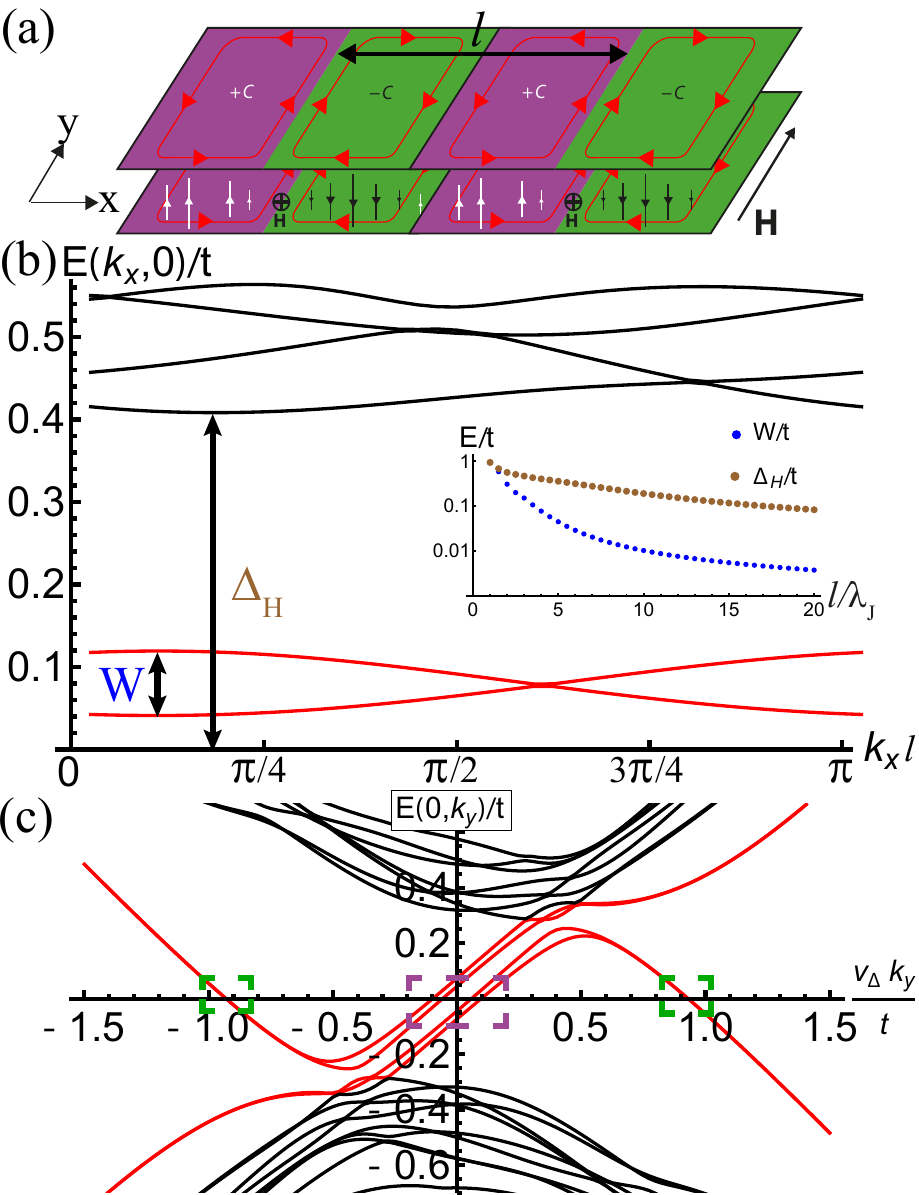}
	\caption{(a) In-plane magnetic field generates a periodic (along $x$) Josephson vortex lattice leading to a pattern of gapped domains with edge modes forming bands. (b) and (c) Quasiparticle energies along $x$ (a) and $y$ (b) direction in the presence of an in-plane field parallel to $k_\perp$. A narrow band (red) is formed within the spectral gap (a). Inset shows the bandwidth and gap as a function of vortex lattice period $l$, main panel is for $l=4\lambda_J$. Dispersion along $k_y$ (b) for $l=6\lambda_J$ shows the low-energy bands crossing zero with a well-defined chirality. Contributions to the LDOS from the marked momenta are shown in Fig.~\ref{fig:ldos}(b). In (a,b), $\alpha=0.5, \frac{v_F}{t \lambda_J} = 0.5$.}
	\label{fig:bands}
\end{figure}

\begin{figure}[h!]
	\includegraphics[width=\linewidth]{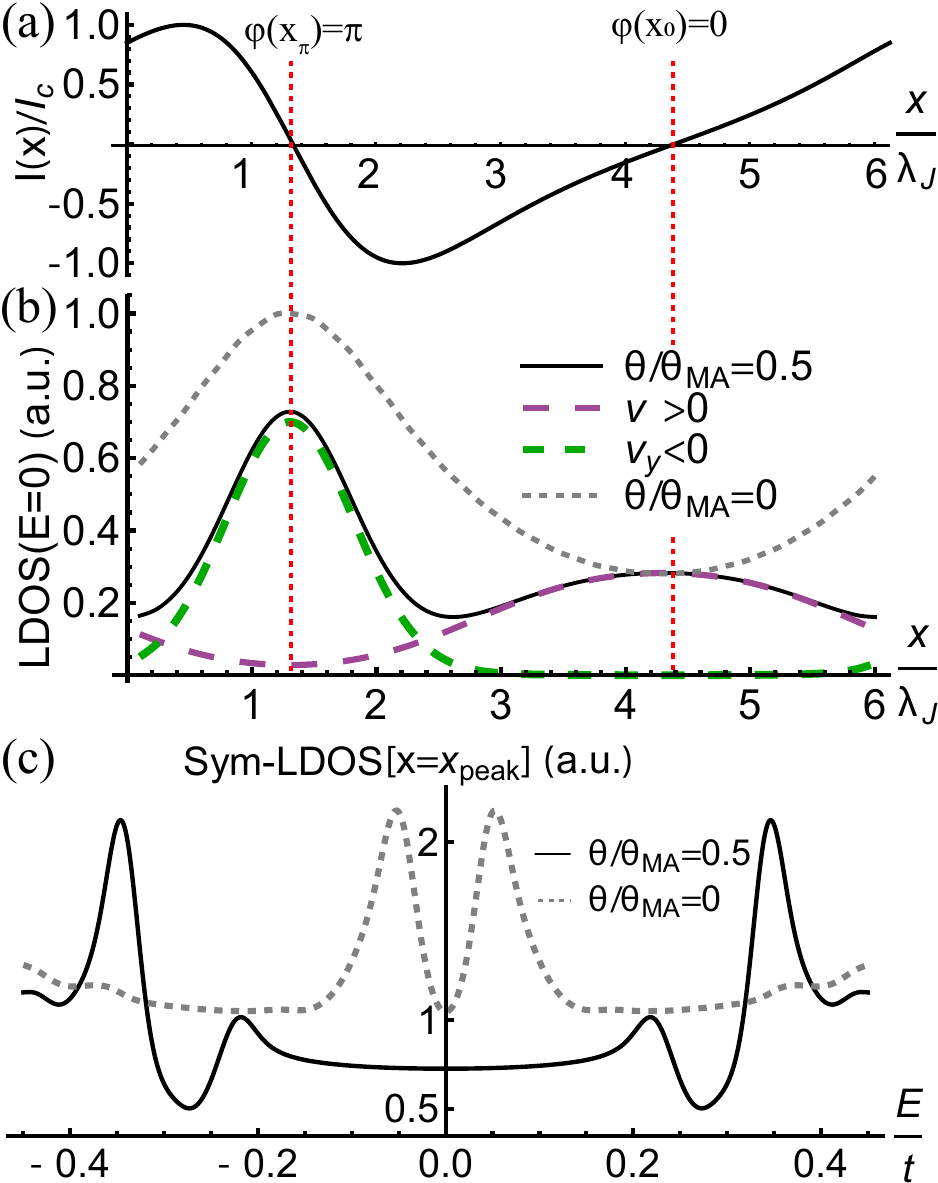}
	\caption{(a) Interlayer current in the presence of an in-plane field (Fig.~\ref{fig:cart} (d)) over one period $l=6 \lambda_J$ of the Josephson vortex lattice with period. (b) Local density of states (LDOS) at zero energy at the top layer. Green and purple lines show contributions of low-energy modes with different chirality (see Fig.~\ref{fig:bands}(b)). (c) Symmetrized energy dependence of LDOS at the peak position: for finite $\theta$ LDOS is constant below within the intra-domain gap.} 
	\label{fig:ldos}
\end{figure}

In this work, we demonstrate that twisted bilayers of two-dimensional nodal superconductors (TBSCs) (Fig.~\ref{fig:cart})
realize topological phases on application of current or magnetic field at any nonzero twist angle. An interlayer Josephson current opens a topological gap that is maximal at a value of the twist angle much smaller than the one required for spontaneous time-reversal breaking \cite{kuboki1996,sigrist1998,can2020hightemperature} and is gradually suppressed for large twist angles (Fig.~\ref{fig:gap}(a)). We also show that an in-plane magnetic field creates a network of topological domains with alternating Chern numbers and chiral edge modes between them (Fig.\ref{fig:bands}(a)). We demonstrate the fingerprints of these tunable topological phases in thermal Hall effect (Fig.~\ref{fig:gap}(b)) and local density of states (Fig.~\ref{fig:ldos}).

{\it Low-energy model of a twisted nodal superconductor bilayer:} We first construct a momentum space low-energy model of a TBSC (illustrated in Fig.~\ref{fig:cart}(a)). The relative rotation of the layers is reflected in the single-particle dispersion $\varepsilon({\bf K})\tau_3\to \varepsilon({\bf K}^{\mp\theta/2})\tau_3$ and pairing $\Delta({\bf K})\hat{\Delta}\to\Delta({\bf K}^{\mp\theta/2})\hat{\Delta}$ terms (where ${\bf K}^{\theta}$ denotes ${\bf K}$ rotated by $\theta$, and $\tau_i$ are Pauli matrices in Gor'kov-Nambu space). Here we will focus on the singlet case $\hat{\Delta}=\tau_1$
\footnote{In the accompanying Article we provide additional derivation details (including the current-phase relation and effects of the rotation of ${\bf k}$ by $\theta$, a generalization to triplet pairing) and study the correlated phases near magic angle.}.  At low twist angles the twist can be approximated in the vicinity of the nodes by a momentum shift $\tilde{{\bf K}}\approx  {\bf k}^{\theta} + [\hat{z}\times {\bf K}_N] \theta\equiv {\bf k}^{\theta} + {\bf Q}_N$. 
Assuming that the gap nodes are not in proximity to the Brillouin zone boundary and that the tunneling decays fast outside the first Brillouin zone \cite{Note1}, the interlayer tunneling amplitude can be taken as a constant ($t$) between overlapping momenta in Fig.~\ref{fig:cart}(a). This implies that the quasiparticles near ${\bf K}_N$ in one layer can tunnel to a vicinity of only a single corresponding node $\tilde{{\bf K}}_N$ in the other layer (Fig.~\ref{fig:cart}(b)). Such pairs of nodes stemming from the two layers form approximately independent ``valleys.''

At the same time, the setup in Fig.~\ref{fig:cart}(a) constitutes a Josephson junction for weak tunneling. Application of a current lower than a critical one ($I_c$) between the layers (Fig.~\ref{fig:cart}(b)) therefore creates a phase difference between the order parameters of two layers $\Delta_1 \to \Delta_1 e^{i\varphi/2},\Delta_2 \to \Delta_2 e^{-i\varphi/2}$. The current-phase relation \cite{golubov2004} at low twist angles takes the form  $I(\varphi)\approx I_c\sin(\varphi)$ \cite{Note1}. We proceed by neglecting rotation of ${\bf k}$, which is appropriate for a circular Fermi surface \cite{Note1} and does not affect qualitative results (see below). The low-energy Hamiltonian of TBSC takes the form $H({\bf k},\varphi) = H_1({\bf k},\varphi)+H_2({\bf k},\varphi)$,
\begin{equation}
    \begin{gathered}
     H_1({\bf k},\varphi)=
    v_F k_\parallel \tau_3+v_\Delta k_\perp \cos(\varphi/2) \tau_1
    \\
    - \alpha t\cos(\varphi/2) \tau_1 \sigma_3 
    +t \tau_3 \sigma_1
    \\
H_2({\bf k},\varphi)=
-v_\Delta k_\perp \sin(\varphi/2)\tau_2\sigma_3
+\alpha t\sin(\varphi/2)\tau_2,
    \end{gathered}
    \label{eq:hamcur}
\end{equation}
where $v_F, v_\Delta$ are the Fermi and gap velocities ($\Delta(K)\approx v_\Delta k_\perp$), $k_{\parallel}(k_\perp)$ are momenta along ${\bf v}_F$ (${\bf v}_\Delta$), $\sigma_i$ are Pauli matrices in layer space and $\alpha=\frac{v_\Delta \theta K_N}{2 t}$. Without current ($\varphi=0$) $H_2({\bf k},\varphi)$ vanishes, while $H_1({\bf k},\varphi)$ has a gapless spectrum \cite{Note1}. For $\varphi\neq0$ a finite spectral gap opens
\begin{equation}
\Delta_J(\varphi)=
\begin{cases}
2  |t \alpha \sin \varphi/2|, &|\alpha| < \cos\varphi/2 \\
\frac{|t \sin \varphi|}{\sqrt{\alpha^2+\sin^2\frac{\varphi}{2}}}
, &|\alpha|> \cos\varphi/2,
\end{cases}
\label{eq:curgap}
\end{equation}
for any $\alpha\neq0$ (i.e. $\theta\neq0$). The gap vanishes for zero interlayer current $I(\varphi)$, i.e. for $\varphi=0,\pi$. In Fig.~\ref{fig:gap}(a), we present the maximal value of the current-induced gap $\Delta_J(\varphi=\varphi_\mathrm{Max})$ (for $\varphi$ between $0$ and $\pi/2$ corresponding to the stable supercurrent branch) as a function of the twist angle. The maximal gap value is equal to $t$ and is reached at $\theta=\theta_\mathrm{MA}/\sqrt{2}$, where $\theta_\mathrm{MA} = \frac{2 t}{v_\Delta K_N}$. To assess the influence of non-circular Fermi surface geometry on the gap we also calculate the spectral gap for a tight-binding Fermi surface appropriate for Bi$_2$Sr$_2$CaCu$_2$O$_{8+ y}$ \cite{markiewicz2005} \footnote{see Supplementary Materials} (Fig.~\ref{fig:gap}(a), red dots).
The circular Fermi surface approximation (dashed line) is in excellent quantitative agreement at $\theta\ll\theta_\mathrm{MA}$. At larger $\theta$, the result can be well-captured by expanding $k^{\pm \theta/2}$ to the lowest order in $\theta$ (solid line) \cite{Note2}. One observes that the gap does not close as a function of twist angle and has an appreciable value for a range of twist angles.

However, for $\theta\gg\theta_\mathrm{MA}$ the gap is strongly suppressed. Note that for the particular case of Bi$_2$Sr$_2$CaCu$_2$O$_{8+y}$, ab-initio estimates suggest $t\approx 1$ meV \cite{markiewicz2005,Note1,Note2} leading to $\theta_\mathrm{MA}\approx2.8^\circ$, which suggests that already at $\theta\approx15^\circ$ the gap would be below $0.025$ meV. In our calculation, we also included the $\propto\cos(2\theta)$ dependence of the interlayer tunneling \cite{Note2} due to the d-wave symmetry of Cu orbitals \cite{song2022}. In the clean case, it vanishes for $\theta$ close to $45^\circ$, where a spontaneous generation of the phase difference was predicted \cite{can2020hightemperature,volkov_jos}. This suggests that the value of the topological gap at low twist angles will be more than order of magnitude larger than in the vicinity of $45^\circ$.

On a qualitative level, the opening of a spectral gap at the nodes in TBSC can be understood to result from a simultaneous breaking of two symmetries: the mirror symmetry of the bilayer (by the twist) and time-reversal symmetry (by the current). Taking the example of a $d_{x^2-y^2}$ SC (relevant for a number of unconventional superconductors \cite{stewart2017}), the breaking of these symmetries allows a mixing of the $d_{x^2-y^2}$ and the $d_{xy}$ order parameters with a relative phase between them, i.e. $d_{x^2-y^2}\to  d_{x^2-y^2}+ e^{i\Phi_{xy}}d_{xy}$. This argument can be similarly generalized to other unconventional superconducting states, i.e. for a triplet $p_x$ superconductor -- under a twist and an applied interlayer current a $p_x+ip_y$ topological superconductor  emerges \cite{Note1,tummuru2020chiral}. The resulting states in all cases are expected to be topological \cite{Kallin_2016,Ghosh_2020,Note1}.

To study the topological properties of our system, we rely on the simpler model of Eq.~\eqref{eq:hamcur} appropriate for $\theta\ll1$. Let us consider the spectrum near the Dirac points of $H_1({\bf k,\varphi})$ at $\alpha\ll1$, where $H_2({\bf k,\varphi})$ can be considered as a perturbation. As the gap does not close with increasing $\alpha$, the topological characteristics apply to all $\alpha\neq 0$. Around $k_\parallel^{\pm} = \pm\sqrt{1-\alpha^2\cos^2(\varphi/2)}t/v_F,k_\perp=0$, projecting the Hamiltonian  Eq.~\eqref{eq:hamcur} to the zero-energy eigenstates of $H_1({\bf k,\varphi})$ one obtains two identical Dirac Hamiltonians:
\begin{equation}
H_{\mathrm{eff}}
=\tilde {v}_{F}(\varphi) k_\parallel \zeta_3 + \alpha t \sin\left(\frac{\varphi}{2}\right)\zeta_2 
+\tilde {v}_{\Delta}(\varphi)k_\perp\zeta_1,
\label{eq:hamdircur}
\end{equation}
where $\tilde{v}_F(\varphi) = \sqrt{1-(\alpha\cos[\varphi/2])^2};\;\tilde{v}_\Delta(\varphi) = \sqrt{1-(\alpha\cos[\varphi/2])^2}\cos[\varphi/2]$.
The Chern number of a single valley with two gapped Dirac points is then equal to $\pm1$ \cite{bernevigbook}; the expression valid for arbitrary $\varphi$ is:
\begin{equation}
    C = \sgn[ v_\Delta \theta \sin(\varphi)].
    \label{eq:chern}
\end{equation}

Moreover, one can demonstrate that the Chern numbers of different valleys are the same. Consider two adjacent nodes on a single layer's Fermi surface [Fig.~\ref{fig:cart}(a)]. While the Fermi velocity changes smoothly between the two and does not vanish anywhere in between (i.e., $v_F$ does not change sign), $v_\Delta$ has to pass through a zero, leading to $v_\Delta\to-v_\Delta$ and $\alpha\to-\alpha$ (after a coordinate rotation) in Eq.~\eqref{eq:hamcur}. Consequently, at the Dirac points in Eq.~\eqref{eq:hamdircur}, the last two terms change sign. This results in the Chern number of two adjacent valleys being the same. The total Chern number is then given by $C_\mathrm{tot} = N_v C$, where $N_v$ is the number of valleys - equal to the number of nodes in a single layer.

We have proven that the interlayer current transforms nodal TBSCs into a topological state characterized by a $\mathbb{Z}$ topological invariant belonging to the C and D Altland-Zirnbauer symmetry
classes \cite{schnyder2008} for singlet and triplet SCs, respectively. The topological nature of these 
states produces gapless neutral chiral (Majorana for the equal-spin triplet pairing case) modes at the edges of the system [Fig.~\ref{fig:cart}(b)], expected to result in a quantized thermal (and spin, for the singlet case) Hall conductance $\kappa_{xy} = n T (\pi/6) k_B^2/\hbar$ at low temperatures \cite{senthil1999,Kallin_2016}, where $n$ is integer. 

To verify this general prediction, we have calculated the thermal Hall conductivity \cite{Note2} for Eq.~\eqref{eq:hamcur} using the expressions in Refs.~\onlinecite{vafek2001,fujimoto2013,cvetkovic2015}. In Fig.~\ref{fig:gap}(b), we present the thermal Hall conductivity $\kappa_H(T)$ normalized to  $-C T (\pi/6) k_B^2/\hbar$ \cite{Note1}. For all nonzero twist angles, the quantization occurs, albeit at temperatures considerably lower than the gap. The temperature at which $\kappa_H(T)$ becomes appreciable does not strongly depend on $\theta$, and is around $0.3 t$, i.e.~3K using values appropriate for Bi$_2$Sr$_2$CaCu$_{2}$O$_{8+\delta}$.

{\it Topological domains induced by an in-plane field:} We now consider the quasiparticles in TBSC in presence of an in-plane magnetic field instead of a current. Extending the analogy with Josephson junction, one expects the emergence of a periodic modulation of the phase difference and current between layers \cite{baronepaterno,Tinkham}---a lattice of Josephson vortices (Fig.~\ref{fig:bands}(a)). The alternating current pattern along $x$ (Fig.~\ref{fig:ldos}(a)) suggests that quasiparticles should be gapped apart from lines (along $y$) where the current vanishes and current-induced gap $\Delta_J$ changes sign. These lines form domain walls between domains with Chern number equal to $\pm C_\mathrm{tot}$.

To study the dispersion of the quasiparticles in the presence of magnetic field, we obtain the BdG Hamiltonian in real space from Eq.~\eqref{eq:hamcur} by $\varphi\to\varphi({\bf r})$, $f({\bf r}) k_i\to \frac{1}{2}\{(-i\partial_i - e/c A_i({\bf r})),f({\bf r})\}$ \cite{simon1997}, where $\{ \cdot , \cdot \}$ denotes the anticommutator. The form of $\varphi(x)$ is determined by the solution of Josephson equations $\frac{\partial^2 \varphi(x)}{\partial x^2} = \frac{1}{\lambda_J^2} \sin \varphi(x)$ \cite{owen1967} (where $\lambda_J^2=\frac{c |\Phi_0|s'}{8\pi^2 |j_c|\left(2\lambda_{ab}^2\right)}$, $\lambda_{ab}$ being the penetration depth, $\Phi_0 = \frac{\pi \hbar c}{e}$, $j_c$ - the critical  current density and $s'$ -thickness of a single layer) \cite{Note2}. The solution is a periodic function with period $l = \Phi_0/(|H|s)$ \cite{Note2}.

We will now focus on the case $k_\parallel \parallel x, k_\perp \parallel y$, with results for different field orientations being qualitatively similar \cite{Note2}. $k_y$ remains a good quantum number, while $k_x$ is folded into a Brillouin zone $k_x\in(0,2\pi/l)$. In Fig.~\ref{fig:bands}(b) an example of the quasiparticle dispersion along $x$ (note that the bands are additionally folded twofold due to the numerical solution procedure \cite{Note2}). One observes a narrow band inside a gap $\Delta_H$ (there is another one at a negative energy). Inset demonstrates that both the width of the narrow band $W$ 
and the gap $\Delta_H$ scale as a function of lattice period inversely proportional to the magnetic field. The dispersion of the in-gap bands along $y$-direction is shown in red in Fig.~\ref{fig:bands}(c). They cross zero energy and merge with other bands afterwards, reminiscent of the edge states in a topological state.

Indeed, this analogy can be confirmed by analyzing the local density of states (LDOS) at zero energy (Fig.~\ref{fig:ldos}(b)); a quantity that can be measured in scanning tunnel microscopy experiments. We plot the LDOS of one layer of TBSC as in an experiment, only LDOS of the layer closest to the tip will be probed. The position of two peaks in LDOS corresponds exactly to points (Fig.~\ref{fig:ldos}(a)), where the current between the layers vanishes. Furthermore, the contributions of states that have opposite chirality (marked by green and purple lines in Fig.~\ref{fig:bands}(b)) are localized at different positions. This confirms the expectation from Fig.~\ref{fig:bands}(a), that the adjacent domain walls host modes moving with opposite velocity along $y$.

Additional insight can be obtained by analyzing the Hamiltonian in the vicinity of the points where interlayer current vanishes (Fig.~\ref{fig:ldos}(a)). Taking only the two states closest to zero energy, in analogy to Eq.~\eqref{eq:hamdircur} the Hamiltonian can be brought to the form of a Dirac equation in a linear confining potential \cite{Note2}: 
\begin{equation}
    H(x\approx x_0 [x_\pi]) = [\alpha] v_F (-i\partial_x) \zeta_3 
    +[-] v_\Delta k_y \zeta_1 + \frac{\alpha q_{0[\pi]} x t}{2} \zeta_2,    
\end{equation}
where $q_{0,\pi} = \varphi'(x_{0,\pi})$ and the effect of the vector potential has been absorbed into a momentum shift. This Hamiltonian has a localized (in $x$) solution with a linear dispersion along $y$ $E_{0[\pi](k_y)}=[-]v_\Delta k_y$, in agreement with Fig.~\ref{fig:bands}(c). The spatial extent of the corresponding eigenfunctions $\psi_{0,\pi}(x,k_y)$, i.e. $\sqrt{\langle\psi_{0[\pi]}| (x-x_{0[\pi]})^2|\psi_{0[\pi]}\rangle}$ is independent of $k_y$ and equal to $\sqrt{v_F/(\alpha t q_0)}$ around $x_0$ and $\sqrt{\alpha v_F/(tq_\pi)}$ around $x_\pi$. Noting that $q_\pi<q_0$ and $\alpha<1$ $\psi_{\pi}$ should be localized much stronger, as is indeed the case in Fig.~\ref{fig:ldos}(b).

We now compare the results with the case $\theta=0$, where the topological gap vanishes (Eq.~\ref{eq:curgap}). The spatially resolved zero-energy LDOS has only a single peak (Fig.~\ref{fig:ldos}(b)). More importantly, the LDOS exhibits strikingly different energy dependence (Fig.~\ref{fig:ldos}(c)). In Fig.~\ref{fig:ldos}(c), the symmetrized energy dependence of the LDOS at the left peak is shown: for energies within the gap $\Delta_H$, the LDOS is constant for  $\theta/\theta_\mathrm{MA} =0.5$, but not for $\theta=0$, where the spectrum is gapless. This behavior is generic and can be also observed at other positions \cite{Note2}.

{\it Effects of disorder}: Gapped topological states are expected to be robust to weak perturbations~\cite{senthil1999,schnyder2008}. To illustrate this general principle we have analyzed Eq.~\ref{eq:hamdircur} in the presence of a random potential $\langle V_\mathrm{imp}({\bf r}) V_\mathrm{imp}({\bf r}') \rangle = n u_0^2$, $n$ being impurity concentration and $u_0$ - scattering strength, in the Born approximation \cite{Note2}. For $\varphi=0$, at arbitrarily weak disorder strength, density of states becomes nonzero at zero energy \cite{balatsky2006,Note2}. In contrast to that, for a finite $\varphi$, DOS remains zero for weak scattering $n u_0^2 \ll 4\pi v_F v_\Delta/\log(\Delta_0/\alpha t \sin(\varphi/2))$, showing that the topology of the state is robust to weak disorder. For the domain wall state in a magnetic field, scattering between edge modes with different $v_y$ could in be important and open a gap. However, their separation in real space (Fig.~\ref{fig:ldos}(b)) reduces the amplitude of the scattering that is proportional to $| \psi_0(x_\mathrm{imp}) \psi_\pi(x_\mathrm{imp})|^2$, where $x_\mathrm{imp}$ is the position of a point-like impurity. For example, for parameters used in Fig.~\ref{fig:ldos}, one obtains averaging over $x_\mathrm{imp}$ that the scattering rate is reduced by a factor of 3 compared to scattering between plane waves \cite{Note2}.

{\it Conclusion:} To conclude, we have shown that twisted bilayers of nodal superconductors can realize topological superconductivity of the neutral BdG quasiparticles ``on demand'' with present-day experimental techniques and systems. Applying an interlayer current bias opens a topological gap in the system that manifests itself in quantized thermal Hall response [Fig.~\ref{fig:gap}(b)]. The gap value is maximized [Fig.~\ref{fig:gap}(a)] near the ``magic'' value of the twist angle. Similarly, the orbital effect of an in-plane magnetic field creates a network of chiral domains separated by Josephson vortex cores hosting chiral one-dimensional modes [Fig.~\ref{fig:bands}(a)]. With several candidate materials proposed to observe these effects~\cite{Note1}, twisted bilayers of nodal superconductors offer a realistic, tunable platform for topological superconductivity.

\section*{Acknowledgments} We thank Philip Kim for insightful discussions. P.A.V.\ is supported by a Rutgers Center for Material Theory Postdoctoral Fellowship and
J.H.P.\ is partially supported by the Air Force Office of Scientific Research under Grant No.~FA9550-20-1-0136, the NSF CAREER Grant No. DMR-1941569, and the Alfred P. Sloan Foundation through a Sloan Research Fellowship. The Flatiron Institute is a
division of the Simons Foundation.

\bibliography{MASC_Lett}

\newpage

\clearpage
\onecolumngrid
\appendix
\renewcommand{\thefigure}{S\arabic{figure}}
\addtocounter{equation}{-2}
\addtocounter{figure}{-3}
\renewcommand{\theequation}{S\arabic{equation}}
\renewcommand{\thetable}{S\arabic{table}}

\begin{center}
	\textbf{\Large
		Supplemental Material for:\\ Current- and field-induced topology in twisted nodal superconductors}
\end{center}

\tableofcontents

\section{Calculation of the spectral gap}

Here we provide additional details on the calculation of the spectral gap under a current bias, Fig. 1 of the main text.

\subsection{Circular Fermi surface}
For the circular Fermi surface, the maximal spectral gap can be found from Eq. 2 of the main text analytically, by minimizing the function for $0<\varphi<\pi/2$. This yields the following result:
\begin{equation}
	\varphi^{max}
	=
	\begin{cases}
		\pi/2, &|\alpha| < 1/\sqrt{2} \\
		2 \arccos(|\alpha|), & 1/\sqrt{2}<|\alpha| <\sqrt{\frac{\sqrt{5}-1}{2}} \\
		2 \arccos(\sqrt{1 + \alpha^2 - \sqrt{\alpha^4 + \alpha^2}}), & \sqrt{\frac{\sqrt{5}-1}{2}}<|\alpha|
	\end{cases},
	\label{eq:phimax}
\end{equation}
and the gap is:
\begin{equation}
	\Delta_J^{max}(\varphi=\varphi_{max})/|t|
	=
	\begin{cases}
		\sqrt{2} |\alpha|, &|\alpha| < 1/\sqrt{2} \\
		2 |\alpha| \sqrt{1 - \alpha^2}, & 1/\sqrt{2}<|\alpha| <\sqrt{\frac{\sqrt{5}-1}{2}} \\
		2 (\sqrt{1 + \alpha^2} - \alpha), & \sqrt{\frac{\sqrt{5}-1}{2}}<|\alpha|.
	\end{cases}
	\label{eq:deltamax}
\end{equation}

\subsection{Beyond the circular Fermi surface}

\subsubsection{Tight binding model for Bi$_2$Sr$_2$CaCu$_2$O$_{8+ y}$}

We consider the opening of the current-induced gap for a tight-binding dispersion for Bi$_2$Sr$_2$CaCu$_2$O$_{8+ y}$ \cite{markiewicz2005}. Each unit cell contains two CuO$_2$ planes, described by the Hamiltonian:
\begin{equation}
	\begin{gathered}
		H({\bf k},\varphi)
		=
		\begin{pmatrix}
			\varepsilon({\bf k})	& e^{i\varphi}\Delta({\bf k})& t_b({\bf k}) &0\\
			e^{-i\varphi}\Delta({\bf k}) & -\varepsilon({\bf k}) &0&- t_b({\bf k})\\
			t_b({\bf k}) &0 &\varepsilon({\bf k})	& e^{i\varphi}\Delta({\bf k})\\
			0&- t_b({\bf k}) & e^{-i\varphi}\Delta({\bf k}) & -\varepsilon({\bf k})
		\end{pmatrix},
		\\
		\varepsilon({\bf k})=-2t(\cos k_x+ \cos k_y)-4t'\cos k_x \cos k_y - 2 t''(\cos 2 k_x+\cos 2 k_y)-\mu,
		\\
		\Delta({\bf k})= \Delta_0 (\cos k_x - \cos k_y),
		\\
		t_b({\bf k}) =  t_b[(\cos k_x-\cos k_y)^2/4 +a_0]
	\end{gathered}
	\label{eq:tb1L}
\end{equation}
$t=126$ meV, $t'=-36$ meV, $t''=15$ meV, $\mu=-130$ meV, $t_b=30 meV$, $a_0=0.4$ \cite{markiewicz2005} and $\Delta_0= 37.5$ meV \cite{vishik2010njop,fedorov1999}. The twisted bilayer with a phase difference is then described as:
\begin{equation}
	\begin{gathered}
		\begin{pmatrix}
			H({\bf k}^{-\theta/2},\varphi/2) & \hat{t}_{\theta}({\bf k})\\
			\hat{t}^T_{\theta}({\bf k}) & H({\bf k}^{\theta/2},-\varphi/2)
		\end{pmatrix},
		\\
		\hat{t}({\bf k})
		=
		\begin{pmatrix}
			0&0&0&0\\
			0&0&0&0\\
			t_\theta({\bf k})&0&0&0\\
			0&-t_\theta({\bf k})&0&0\\
		\end{pmatrix}
		\\
		\hat{t}_{\theta}({\bf k}) = t_z[(\cos k_x^{\theta/2}-\cos k_y^{\theta/2})(\cos k_x^{-\theta/2}-\cos k_y^{-\theta/2})/4 +a_0 \cos(2\theta)],
	\end{gathered}
	\label{eq:bscco}
\end{equation}
where the symmetry of the Cu $d_{x^2-y^2}$ orbitals have been taken into account \cite{song2022} and $t_z=5$ meV  \cite{markiewicz2005}.

\subsubsection{Low-$\theta$ expansion}

Additionally, we attempt to capture the dependence of the current-induced gap in a tight-binding model by introducing corrections to Eq. (1) of the main text that take into account non-circular Fermi surface geometry. For simplicity, we have neglected the bilayer splitting and considered a two-layer model. To the lowest order in $\theta$ these are given by \footnote{See the accompanying Article for additional details}:
\begin{equation}
	\delta \hat{H}_\theta \approx
	\frac{v^{(2)}_F \theta k_\perp}{2} \tau_3\sigma_3
	- \frac{v^{(2)}_{\Delta} \theta k_\parallel}{2} \cos(\varphi/2)\tau_1\sigma_3
	+\frac{v^{(2)}_{\Delta} \theta k_\parallel}{2} \sin(\varphi/2)\tau_2
	,
	\label{eq:hamkrot}
\end{equation}
where 
\begin{equation}
	v^{(2)}_F = v_F - K_N \frac{\partial^2 \varepsilon({\bf k})}{\partial k_\perp^2}
	;\;
	v^{(2)}_{\Delta}=v_\Delta + K_N \frac{\partial^2 \Delta({\bf k})}{\partial k_\parallel \partial k_\perp}.
	\label{eq:v2def}
\end{equation}
$v^{(2)}_F $ and $v^{(2)}_{\Delta}$ are determined from Eq. \eqref{eq:tb1L} neglecting bilayer splitting. At the same time, as the intercell coupling in Eq. \eqref{eq:bscco} affects only two layers out of four, effective intercell coupling for the bonding/antibonding bands is reduced by two \footnotemark[\value{footnote}]. Therefore, $t$ is taken to be equal to $a_0 t_z/2=1$ meV, ignoring the $\cos(2\theta)$ factor, appropriate for low twist angles. $\theta_{MA}$ estimate in this case is $2.8^\circ$. The values of the current-induced gap were found by minimizing the resulting Hamiltonian eigenvalues numerically. One observes from Fig. 1 in the main text, that this approximation recovers the gap value of the tight binding model well up to $\theta = 5 \theta_{MA}$.

\section{Thermal Hall}

The intrinsic thermal Hall conductivity is calculated from the general formula \cite{cvetkovic2015} (see also\cite{vafek2001,fujimoto2013}):
\begin{equation}
	\begin{gathered}
		\kappa_H(T) = \frac{1}{\hbar T} \int_{-\infty}^{\infty} d\xi\xi^2 \left(-\frac{\partial f(\xi)}{\partial \xi}\right) \tilde{\sigma}_H(\xi),
		\\
		\tilde{\sigma}_H(\xi)=-i \int \frac{d^2 k}{4\pi^2} \sum_{E_m({\bf k})<\xi<E_n({\bf k})}
		\frac{\left \langle m{\bf k}\left|\frac{\partial H}{\partial k_\parallel}\right|n{\bf k}\right\rangle\left\langle n{\bf k}\left|\frac{\partial H}{\partial k_\perp}\right|m{\bf k}\right\rangle - (\parallel\leftrightarrow \perp)}{(E_m({\bf k})-E_n({\bf k}))^2}.
	\end{gathered}
\end{equation}

At low temperatures, the expression above can be reduced to $\kappa_H(T) |_{T\to0}= T C_H \frac{\pi}{6} \frac{k_B^2}{\hbar}$, where $C_H$ is 
\begin{equation}
	C_H=\frac{1}{2\pi i} \sum_n\int d^2 k[\nabla_{\bf k}\times\langle n {\bf k}|\nabla_{\bf k}|n {\bf k}\rangle]_z \theta[-E_n({\bf k})].
\end{equation}
Note that this definition is opposite in sign to that used in quantum Hall effect - compare with (3.9), (4.9)\cite{kohmoto1985} or compare equation after Eq. (20) in \cite{fujimoto2013} and (2.12) in \cite{bernevigbook}.

\section{Orbital effects of in-plane magnetic field}

\subsection{Magnetic field distribution and phase profile}

\begin{figure}[h!]
	\includegraphics[width=\linewidth]{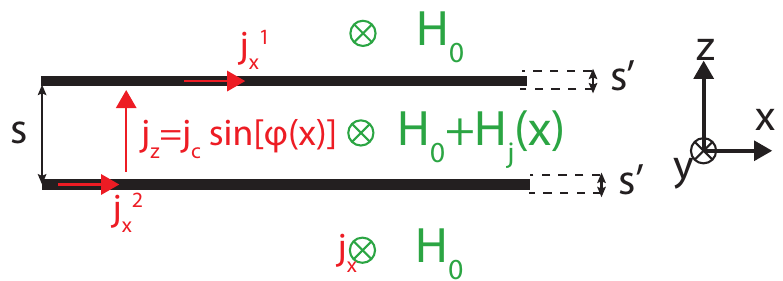}
	\caption{Geometry of TBSC in a parallel magnetic field $H_0$ (Twist between layers is not shown here.). $H_j(x)$ is the screening field of the supercurrents, red arrows mark the current flow directions (for definitions see text). }
	\label{fig:geom}
\end{figure}

In-plane magnetic field leads to the appearance of a nonzero vector potential as well as a position dependent phases of the superconducting order parameter of the two layers $\Phi_{1,2}$ \cite{baronepaterno}. Geometry is presented in Fig. \ref{fig:geom}. We chose the $y$ axis to be along the field and $z$ axis perpendicular to the TBSC plane. 

To find the dependence of the magnetic field and superconducting phases on coordinates, we use Maxwell and London equations (for additional discussion see Ref. \cite{volkov_jos}). Within the superconducting layers of thickness $s'$ one has:
\begin{equation}
	\frac{\partial H}{\partial z} =  - \frac{4\pi}{c} j_x^{1,2} = \frac{\Phi_0}{2\pi\lambda_{ab}^2} 
	\left[\frac{\partial \Phi_{1,2}}{\partial x}(x,z)+\frac{2\pi}{\Phi_0} A_x(x,z)\right],
	\label{eq:sclayers}
\end{equation}
where $\lambda_{ab}$ is the penetration depth and $\Phi_0=\frac{\pi \hbar c}{|e|}$. For BSCCO, the relevant value of $\lambda_{ab}$  is at least $2100$ \AA \cite{enriquez2001}. Outside the TBSC junction region it is equal to the applied field, while inside the junction there is a position-dependent correction denoted $H_j(x)$ due to the supercurrents. Assuming $s'\ll \lambda_{ab}$ we can integrate Eq. \eqref{eq:sclayers} across each layer neglecting the dependence of the r.h.s. on $z$. Subtracting the results for two layers one gets:
\begin{equation}
	H_j(x) =  - \frac{\Phi_0 s'}{2\pi\lambda_{ab}^2} 
	\left[\frac{\partial \Phi_1-\Phi_2}{\partial x}(x)+\frac{2\pi}{\Phi_0} (A_x(x,z_1)- A_x(x,z_2))\right],
	\label{eq:hjx}
\end{equation}
where $z_1$ and $z_2$ are the coordinates of the two layers. Assuming the magnetic field variations to occur at a scale much larger than the interlayer distance $s$ we can further bring this equation to the form:
\begin{equation}
	\begin{gathered}
		H_j(x) =  - \frac{\Phi_0 s'}{4\pi\lambda_{ab}^2} 
		\left[\frac{\partial \Phi_1-\Phi_2}{\partial x}(x)+\frac{2\pi s}{\Phi_0} \frac{\partial A_x}{\partial z}(x)\right]=
		- \frac{\Phi_0 s'}{4\pi\lambda_{ab}^2} 
		\left[\frac{\partial \varphi}{\partial x}(x)+\frac{2\pi s}{\Phi_0} (H_0+H_j(x))\right],
	\end{gathered}
	\label{eq:hjxappr}
\end{equation}
where we introduced the gauge-invariant phase difference across the junction \cite{baronepaterno,glazman1992} (note that $z_1>z_2$):
\begin{equation}
	\varphi(x) = \Phi_1(x)-\Phi_2(x)+\frac{2\pi}{\Phi_0}\int_{z_2}^{z_1} A_z(x) dz,
	\label{eq:phigauge}
\end{equation}
and used $H(x)=H_0+H_j(x)=[ \nabla \times {\bf A}]$. Finally, one can rearrange \eqref{eq:hjxappr} to get the expression for $H_j(x)$:
\begin{equation}
	H_j(x)= -\frac{\frac{\Phi_0}{2\pi} \frac{\partial \varphi}{\partial x}}{\frac{2\lambda_{ab}^2}{s'}+s}
	-\frac{H_0 s}{\frac{2\lambda_{ab}^2}{s'}+s}.
\end{equation}
One can use now Maxwell's equation $\frac{\partial H}{\partial x} = \frac{4 \pi} {c} j_z(x)$ combined with Josephson relation $j_z(x) = -|j_c| \sin\varphi(x)$ to get \cite{owen1967}:
\begin{equation}
	\frac{\partial^2 \varphi(x)}{\partial x^2} = \frac{1}{\lambda_J^2} \sin \varphi(x);
	\;
	\lambda_J^2  =  \frac{c \Phi_0}{8\pi^2 |j_c|\left(\frac{2\lambda_{ab}^2}{s'}+s\right)}.
	\label{eq:josephson}
\end{equation}
As the critical current is mostly determined by the region away from nodes \cite{volkov_jos}, we take $\lambda_J$ as a phenomenological parameter. For $\frac{2\lambda_{ab}^2}{s'} \gg s$, $\lambda_J\approx \gamma\sqrt{\frac{ss'}{2}} $ \cite{glazman1992}, where $\gamma=\lambda_c/\lambda_{ab}$, which is around $1000$ in BSCCO  \cite{latyshev1996}. Therefore, taking $s = 1.5$ nm and estimating $s'$ to be of the order $0.3$ nm (the thickness of two CuO$_2$ layers within the unit cell from TEM image \cite{frank_exp}), one can expect the Josephson length in twisted BSCCO at low twist angles to be of the order of 0.5 $\mu$m.

The solution of the equations \eqref{eq:josephson} are given by Jacobi elliptic functions with two parameters, $k$ and $x_0$ that are determined by the boundary conditions \cite{owen1967}. Here we will assume an infinitely long  $L\gg \lambda_J$ system and fix these parameters in a different way. We assume that the total current through the junction vanishes; this is satisfied by a periodic solution of Eq. \eqref{eq:josephson}
\begin{equation}
	\begin{gathered}
		\frac{\partial \varphi(x)}{\partial x} =
		\frac{2}{k \lambda_J} {\rm dn} \left(\left. \frac{x-x_0}{k \lambda_J} \right| k^2\right);
		\\
		\sin[\varphi(x)/2] = {\rm cn} \left(\left. \frac{x-x_0}{k \lambda_J} \right| k^2\right);
		\\
		\cos[\varphi(x)/2] = -{\rm sn} \left(\left. \frac{x-x_0}{k \lambda_J} \right| k^2\right),
	\end{gathered}
	\label{eq:jossol}
\end{equation}
with the period $l$ given by:
\begin{equation}
	l = 2 k \lambda_J K(k^2),
\end{equation}
where using $0<k<1$ any period $l\in(0,\infty)$ can be realized. One observes that the parameter $x_0$ simply shifts the solution and therefore can be taken to have an arbitrary value without loss of generality. We take $x_0$ such that $\left.-\frac{\Phi_0}{2\pi s} \frac{\partial \varphi}{\partial x}\right|_{x=0,L} = H_0$, which ensures good convergence of the quasiparticle band structure. Specifying in addition the period $l$ fully determines $\varphi(x),H(x)$. In particular, we take the period determined by the flux quantization condition:
\begin{equation}
	H_0 s l = -\Phi_0.
	\label{eq:fluxquant}
\end{equation}
We now check using Eq. \eqref{eq:hjxappr} that the above can indeed be satisfied. Integrating both parts over $x$ over a period $l$ and using $\int_0^{2K(k^2)} {\rm dn} (a|k^2) da = \pi$ we find that $\int_0^l dx H_j(x) = 0$. Therefore, in a finite system with size $L=n l$ the boundary conditions $H(0)=H(L)=H_0$ can be always satisfied by choosing $x_0$ such that $H_j(0)=0$.

Furthermore, we can provide an estimate for ${\rm max} |H_j|(x)$ and show that it is negligible compared to $H_0$. Using $|{\rm dn} (a|k^2)|\leq 1$ for $k\leq1$ one obtains $\left|\frac{\partial \varphi(x)}{\partial x}\right|< \frac{2}{k \lambda_J}$. To obtain an upper estimate for $k$ we use $k K(k^2)< \frac{\pi k/2}{1-k}$ resulting in $k>\frac{1}{\frac{\pi \lambda_J H_0 s}{\Phi_0} +1}$
\begin{equation}
	|H_j|(x) \leq \frac{s' \Phi_0}{2\pi \lambda_{ab}^2 \lambda_J}+\frac{ss'}{\lambda_{ab}^2} |H_0|.
\end{equation}
For $s,s'\ll\lambda_{ab}$ the second term is negligible compared to $H_0$, whereas the first term is such only for $|H_0|\gg\frac{s' \Phi_0}{2\pi \lambda_{ab}^2 \lambda_J}$. However, taking the lower estimate of $\lambda_{ab}, \lambda_J$ to be of order $0.1$ microns with $s'$ (which should be less then the unit cell thickness) less then a nanometer this condition implies fields larger then $10^{-3}$ T which can be easily satisfied in modern experiments.

With two length parameters ($l$ and $\lambda_J$), the solution \eqref{eq:jossol} crosses over between two limits: for $l\gg\lambda_J$ the solution is a periodic arrangement of well separated Josephson vortices of size $\lambda_J$, while for $l\ll\lambda_J$ the current distribution is almost sinusoidal, so that $\varphi(x)\approx - 2 \pi H_0 s x/\Phi_0$. The crossover from the first to the second region occurs on increasing $H_0$, the relevant value for BSCCO where $l=\lambda_J$ is (using the estimates above) around $2.7$ T. Strictly speaking, the field can not penetrate the junction below a critical value; this has been found, however, to be of the order few Oe in bulk BSCCO \cite{enriquez2001} and is not expected to be much higher in the bilayer, as the decrease in Josephson length is rather moderate of the order $\sqrt{s'/s}\approx 0.5$.

\subsection{Quasiparticle Hamiltonian}

We now move on to the derivation of the quasiparticle Hamiltonian. We pick the gauge where $A_{y,z} = 0;\; A_x = z H_y(x)$. The phases of the individual order parameters (see Eq. \eqref{eq:phigauge}) can then be simply taken as $\Phi_1(x) = \varphi(x)/2$ and $\Phi_2(x) = -\varphi(x)/2$. Note that the amplitude of the order parameter is unaffected by the in-plane field in a Josephson junction, such as the TBSC.

In the BdG Hamiltonian, a position- and momentum-dependent superconducting order parameter is described as follows. The pairing term in the Hamiltonian takes the form $\Phi_{1,2}^\dagger({\bf r})
[\Delta_{1,2}({\bf r},{\bf r}') \hat{\Delta}]
\Phi_{1,2}^\dagger({\bf r}')$. In the absence of magnetic field, the system is translationally invariant and the pairing field does depend only on ${\bf r}-{\bf r}'$. Using the center of mass coordinates $\Delta_{1,2}([{\bf r}+{\bf r}']/2,{\bf r}-{\bf r}') $ and taking the Fourier transform with respect to the difference, one gets $\Delta_{1,2}([{\bf r}+{\bf r}']/2,{\bf k})$, which is equal to (near the nodes) $v_{\Delta}([{\bf r}+{\bf r}']/2) (k_\perp \mp Q/2)$.

Fourier transforming the expression back to real space with coordinates ${\bf r},{\bf r'}$ \cite{simon1997}, one gets $\Delta_{1,2}({\bf r},{\bf r}') \to \delta({\bf r} - {\bf r}') \frac{1}{2} \{v_\Delta({\bf r}) , [-i\hbar\overrightarrow{\partial}_\perp\mp Q/2]\}$, where $\{...\}$ is the anticommutator \cite{simon1997} and $v_{\Delta}({\bf r}) = v_{\Delta} \cdot e^{\pm i\varphi({\bf r})/2}$.

The magnetic field is introduced with the Peierls substitution ($-i \nabla \to -i \nabla -\frac{e}{c} {\bf A}({\bf r})$). In the gauge $A_{y,z} = 0;\; A_x = z H_y(x)$ one gets $-i\partial_x\to-i\partial_x-\frac{e}{c} \frac{s H(x)}{2} \sigma_3$. 

In the general case, the $k_\parallel$ and $k_\perp$ axis are at an angle $r$ to the $x,y$ axes [$n_\parallel = (\cos r , -\sin r);\;n_\perp = (\sin r , \cos r)$]. As translational invariance along $y$ is intact, one can use the ansatz $\Psi(x,y) = \psi_{k_y}(x) e^{i k_y y}$. The Hamiltonian now takes the general form:
\begin{equation}
	\begin{gathered}
		H(x,k_y)=
		v_F\left[\cos(r) \left(-i \hbar \partial_x - \frac{e}{c} \frac{s H(x)}{2} \sigma_3 \right) +\sin(r) k_y\right]
		\tau_3 
		+  
		\\
		+
		v_\Delta
		\left[
		-\frac{\sin(r)}{2}\{-i \hbar \partial_x - \frac{e}{c} \frac{s H(x)}{2} \sigma_3, \cos(\varphi(x)/2)\} 
		+ \hbar \cos(r)  \cos(\varphi(x)/2)  k_y 
		\right]\hat{\Delta}
		\\
		- \alpha t\cos(\varphi(x)/2)\hat{\Delta} \sigma_3
		+t \tau_3 \sigma_1
		\\
		+v_\Delta
		\left[
		-\frac{\sin(r)}{2}\{-i \hbar \partial_x - \frac{e}{c} \frac{s H(x)}{2} \sigma_3, \sin(\varphi(x)/2)\} 
		+ \hbar \cos(r)  \sin(\varphi(x)/2)  k_y 
		\right]
		i\tau_3 \hat{\Delta} \sigma_3-
		\\
		-\alpha t\sin(\varphi(x)/2)i\tau_3 \hat{\Delta}.
	\end{gathered}
\end{equation}

For a singlet superconductor  ($\hat{\Delta} = \tau_1$) we further simplify the notation, normalizing the Hamiltonian by $t$:
\begin{equation}
	\begin{gathered}
		H(x,k_y)/t=
		\tilde{v}_F\left[\cos(r) \left(-i \hbar \partial_{\tilde{x}} - h(x) \sigma_3 \right) +\sin(r) \tilde{k}_y\right]
		\tau_3 
		+  
		\\
		+
		\tilde{v}_\Delta
		\left[
		-\frac{\sin(r)}{2}\{-i \hbar \partial_{\tilde{x}} - h(x) \sigma_3, \cos(\varphi(x)/2)\} 
		+ \cos(r)  \cos(\varphi(x)/2)  \tilde{k}_y 
		\right]\tau_1
		\\
		- \alpha \cos(\varphi(x)/2) \tau_1\sigma_3
		+ \tau_3 \sigma_1
		\\
		-\tilde{v}_\Delta
		\left[
		-\frac{\sin(r)}{2}\{-i \hbar \partial_{\tilde{x}} - h(x) \sigma_3, \sin(\varphi(x)/2)\} 
		+ \cos(r)  \sin(\varphi(x)/2)  \tilde{k}_y 
		\right]
		\tau_2 \sigma_3+
		\\
		+\alpha \sin(\varphi(x)/2) \tau_2,
	\end{gathered}
	\label{eq:bdg_singlet}
\end{equation}
where $\tilde{x} = x/\lambda_J,\;\tilde{k_y} =\lambda_J k_y,\;\tilde{v}_F = \frac{\hbar v_F}{t\lambda_J},\;\tilde{v}_F = \frac{\hbar v_\Delta}{t\lambda_J},\; h(x) = \frac{e s \lambda_J H(x)}{2 \hbar c}$. From the above considerations (see Eq. \eqref{eq:fluxquant}), for TBSC with $s'\ll\lambda_{ab}$
\begin{equation}
	h(x) = \frac{e s \lambda_J H(x)}{2 \hbar c} \approx -\frac{\pi \lambda_J H_0 s}{2 \Phi_0} = \frac{\pi\lambda_J}{2 l}.
\end{equation}

In the main text, we focus on the case $k_\parallel \parallel x, k_\perp \parallel y$ ($r=0$):
\begin{equation}
	\begin{gathered}
		H(x,k_y)/t =|_{(r=0)} \;\;\;	\tilde{v}_F \left(-i \hbar \partial_{\tilde{x}} - h \sigma_3 \right) \tau_3 
		+  
		\tilde{v}_\Delta \tilde{k}_y \cos(\varphi(x)/2) \tau_1 
		- \alpha \cos(\varphi(x)/2)\tau_1 \sigma_3
		\\
		+ \tau_3 \sigma_1
		-\tilde{v}_\Delta \tilde{k}_y \sin(\varphi(x)/2)  \tau_2 \sigma_3 
		+\alpha \sin(\varphi(x)/2) \tau_2.
	\end{gathered}
	\label{eq:bdg_r=0}
\end{equation}

Eigenvectors take the form of a 4-spinor $ \psi_{n,k_y}(x,y) =[u^{top}_{n,k_y}(x),v^{top}_{n,k_y}(x),u^{bot}_{n,k_y}(x),v^{bot}_{n,k_y}(x)]$. To find the eigenfunctions and eigenvalues numerically, we use the {\it NDEigensystem} routine in {\it Wolfram Mathematica}. Due to discrete step used in calculations, unphysical solutions have been found to appear, characterized by extremely noisy eigenfunctions. We get rid of them by restricting the eigenfunction derivative value, i.e. $\sqrt{|u^{top'}_{n,k_y}(l)|^2+|v^{top'}_{n,k_y}(l)|^2+|u^{bot'}_{n,k_y}(l)|^2+|v^{bot'}_{n,k_y}(l)|^2}<C$ (values of $C$ listed below). Due to presence of half-periodic terms in the Hamiltonian, optimal convergence is achieved for a doubled unit cell, where all terms (note, e.g., $\sin(\varphi/2)$) are strictly periodic. Note that this procedure does not involve any physical assumptions and simply corresponds to formally folding the actual bands onto a twice smaller Brillouin zone, without changing the eigenvalues.

Using the eigenfunctions obtained, one can also evaluate the local density of states, that can be observed in STM experiments. Due to the layered structure of the system, we assume also that the tunneling is restricted to one (top) of the layers only. The density of states is then given by \cite{gygi1991,suematsu2004,halterman2005}:
\begin{equation}
	N(x,y,E) = \sum_{n,k_y}|u^{top}_{n,k_y}(x)|^2 \delta(E-E_{n,k_y}) + |v^{top}_{n,k_y}(x)|^2 \delta(E+E_{n,k_y}),
\end{equation}
where for numerical calculation we used a gaussian level smearing $\delta(E-E_{n,k_y}) \to e^{-(E-E_{n,k_y})^2/\sigma^2}/(\sqrt{\pi}\sigma)$ ($\sigma$ values given below). 

Note that LDOS in a single layer does not have to be a symmetric function of $E$. However, the total LDOS of the two layers can be shown to be a symmetric function of $E$. It follows from $(\sigma_2 C_2)^{-1} H(x,k_y) \sigma_2 C_2 = -H(x,k_y)$ for Eq. \eqref{eq:bdg_singlet}, where $C_2$ is defined with center at the point $\varphi(x) \mod 2\pi=0$: $C_2: x\to-x,\; k_y\to-k_y,\; \sin(\varphi(x)/2) \to - \sin(\varphi(x)/2),\; \cos(\varphi(x)/2) \to  \cos(\varphi(x)/2)$. This symmetry involves an interchange between layers, which are not identical due to the phase difference. Note that the Hamiltonian \eqref{eq:bdg_singlet} does not possess the usual particle-hole symmetry $(K \tau_2)^{-1} H K \tau_2 = -H$ due to $\varphi(x)$ terms.

Interestingly, for the particular case $r=0$ there is an additional symmetry. For a $d$-wave superconductor with $C_2$ (${\bf k}\to-{\bf k}$) symmetry, pairs of $C_2$-related nodes must exist with $v_F\to -v_F, v_\Delta\to -v_\Delta$ and $(K_N\to -K_N)$, such that  $(\alpha\to \alpha)$ in \eqref{eq:bdg_r=0}. Applying a transformation $H(x,k_y) \to (\tau_3 \sigma_1 I_x)^{-1} H(x,k_y) \tau_3 \sigma_1 I_x$ recovers the Hamiltonian of the initial node. Therefore, the sum of LDOS of two $C_2$-related nodes is equal to the LDOS of two layers for one node and therefore is an even function of $E$.

Below we consider concrete examples and present additional details. The parameters used for all calculations are $\tilde{v}_F=0.5,\;\tilde{v}_\Delta=0.05$. In particular, we consider the following cases.

\newpage
\subsection{Numerical calculation results}

\subsubsection{Details for $l=6\lambda_J;\; r=0; \; \alpha=0.5,0$}

\begin{figure}[h!]
	\centering
	\begin{subfigure}{.5\textwidth}
		\centering
		\includegraphics[width=\linewidth]{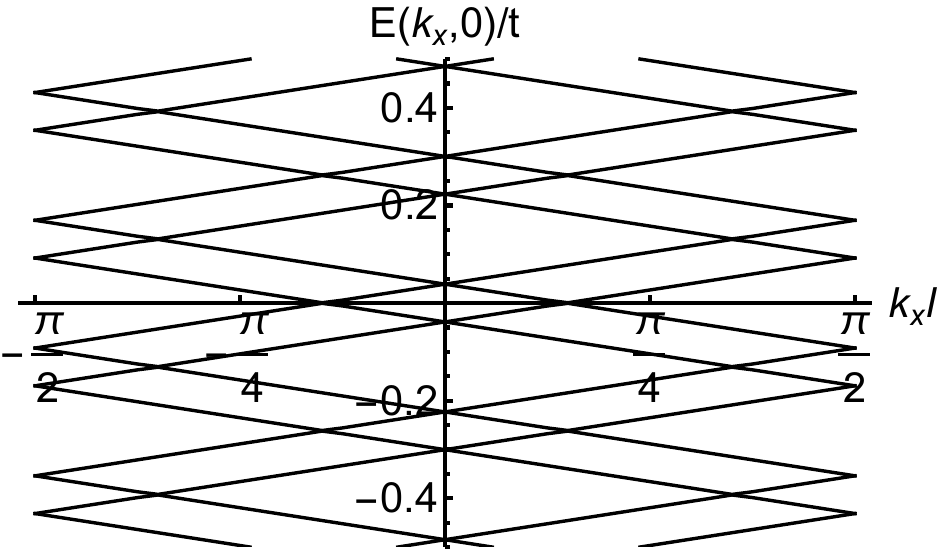}
		\caption{Quasiparticle dispersion along $k_x$, $\theta=0$.}
		\label{fig:bandst0x}
	\end{subfigure}%
	\begin{subfigure}{.5\textwidth}
		\centering
		\includegraphics[width=\linewidth]{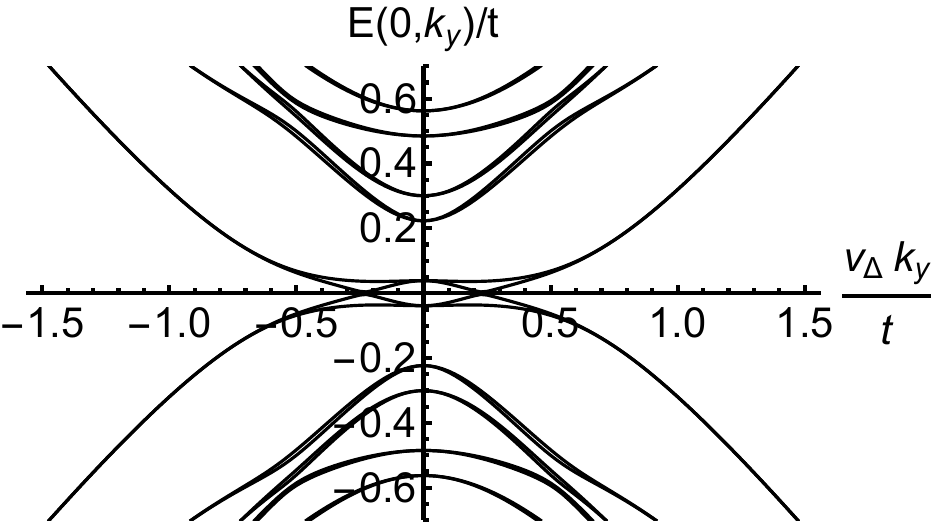}
		\caption{Quasiparticle dispersion along $k_y$, $\theta=0$.}
		\label{fig:bandst0y}
	\end{subfigure}
	\caption{Quasiparticle band structure in magnetic field for $\theta=0$ along (a) $k_x$ and (b) $k_y$. Unlike finite twist $\theta=0.5 \theta_{MA}$ (see main text), it is evident from (b) that the band structure does not have a clear defined gap and in-gap states - all bands are connected.}
	\label{fig:bandst0}
\end{figure}

{\it Band structure:} In Fig. \ref{fig:bandst0} we show the quasiparticle dispersion for the zero-twist case.

\begin{figure}[h!]
	\centering
	\begin{subfigure}{.45\textwidth}
		\centering
		\includegraphics[width=\linewidth]{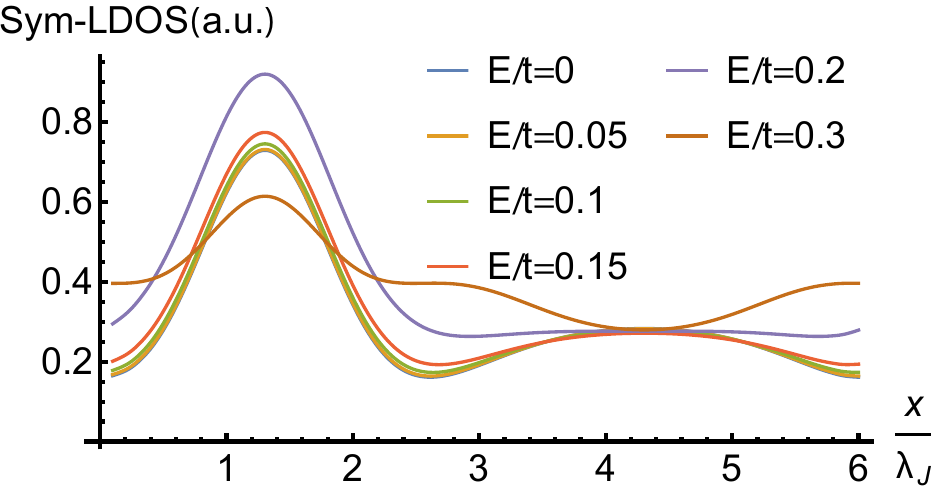}
		\caption{Symmetrized LDOS as a function of real-space coordinate and energy for $\theta=0.5\theta_{MA}$.}
	\end{subfigure}%
	\hspace{4pt}
	\begin{subfigure}{.45\textwidth}
		\centering
		\includegraphics[width=\linewidth]{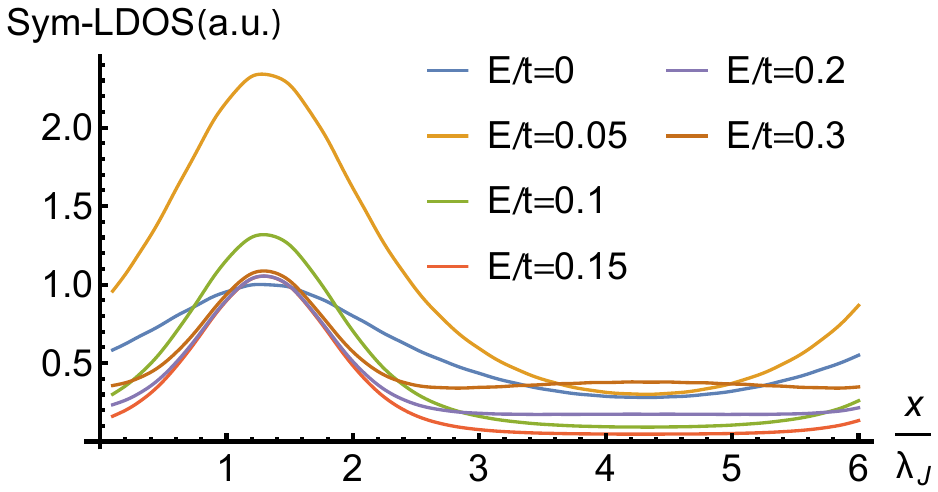}
		\caption{Symmetrized LDOS as a function of real-space coordinate and energy for $\theta=0$.}
	\end{subfigure}
	\caption{Additional details for the LDOS (see Fig. 4 of the main text). LDOS curves almost does not change for a range of energies around $E=0$ for $\theta=0.5\theta_{MA}$, indicating that most of the bands are gapped.}
	\label{fig:L=6ldos}
\end{figure}

{\it LDOS:} In Fig. \ref{fig:L=6ldos} we present the full position-dependence of energy-symmetrized LDOS at several energies for $\theta=0,0.5$.

\begin{figure}[h!]
	\centering
	\includegraphics[width=0.5\linewidth]{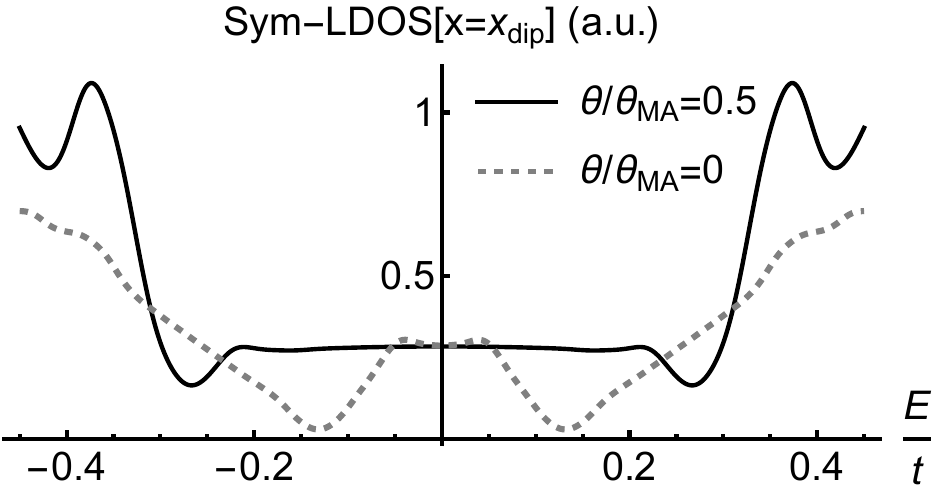}
	\caption{Symmetrized LDOS near the local LDOS dip for $\theta=0$ (or second LDOS peak for $\theta\neq 0$) $x/L\approx 4.305$. A gap of about $0.4t$ is evident for $\theta=0.5\theta_{MA}$. LDOS almost does not change for a range of energies around $E=0$, indicating that most of the bands are gapped.}
	\label{fig:L=6ldosdip}
\end{figure}

In addition, in Fig. \ref{fig:L=6ldosdip} we the symmetrized energy dependence of LDOS at $x/L\approx 4.305$.

{\it Calculation details:} "MaxCellMeasure" of $0.1$, a cutoff value $C=1$ for the eigenfunction derivative at $x=l$, level smearing for LDOS calculation$\sigma/t= 0.015$

\newpage
\subsubsection{Results for $l=6\lambda_J$, $r=\pi/2, \theta/\theta_{MA}=0,0.5$}

We now demonstrate that the qualitative features of spectrum and LDOS discussed in main text do not depend on the direction of the in-plane magnetic field. For $r=\pi/2$, $k_\parallel$ is along the $y$ axis. 

\begin{figure}[h!]
	\centering
	\begin{subfigure}{.5\textwidth}
		\centering
		\includegraphics[width=\linewidth]{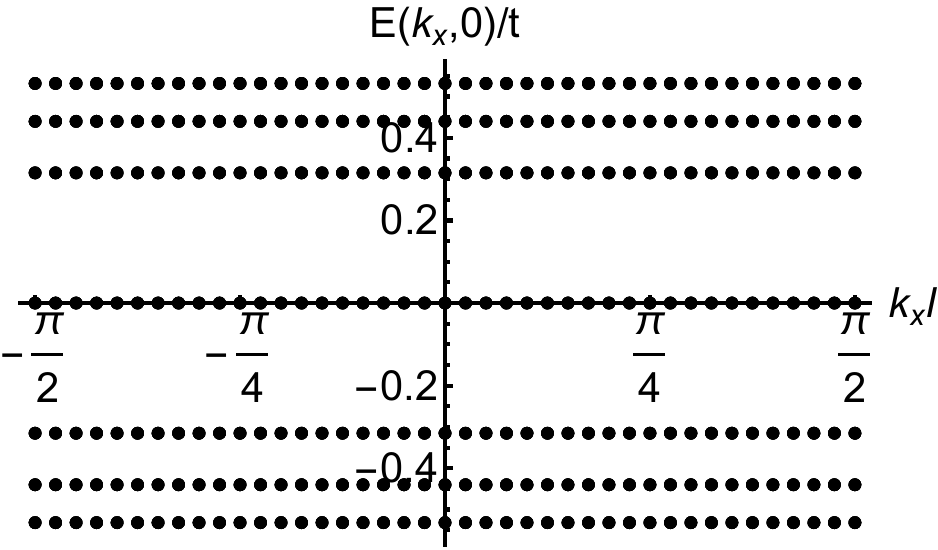}
		\caption{Quasiparticle dispersion along $k_x$, $\theta=0$.}
		\label{fig:bandsr90t0x}
	\end{subfigure}%
	\begin{subfigure}{.5\textwidth}
		\centering
		\includegraphics[width=\linewidth]{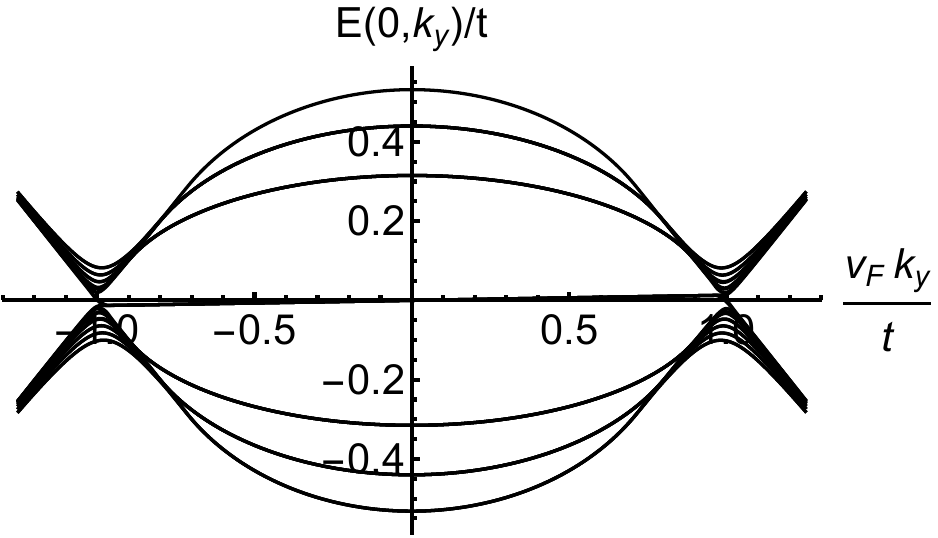}
		\caption{Quasiparticle dispersion along $k_y$, $\theta=0$.}
		\label{fig:bandsr90t0y}
	\end{subfigure}
	\caption{Quasiparticle band structure in magnetic field for $\theta=0$ along (a) $k_x$ and (b) $k_y$ (only few lowest-energy bands are shown). Unlike finite twist $\theta=0.5 \theta_{MA}$ (see main text), it is evident from (b) that the band structure does not have a clear defined gap and in-gap states - all bands are connected.}
	\label{fig:bandsr90t0}
\end{figure}

\begin{figure}[h!]
	\centering
	\includegraphics[width=0.5\linewidth]{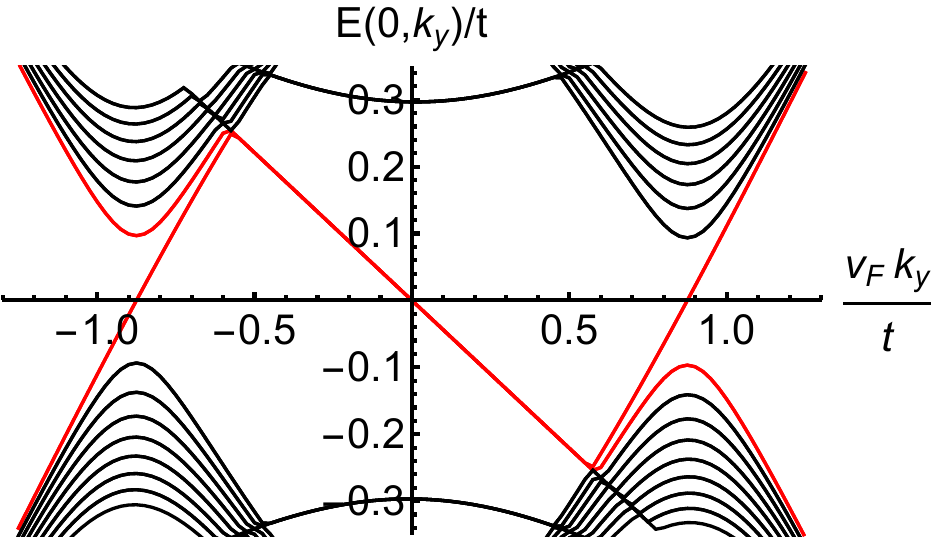}
	\caption{Quasiparticle band structure in magnetic field for $\theta=0.5\theta_{MA}$ along $k_y$ (dispersion along $k_x$ is negligible at low energies). A gap of around $0.2 t$ is found to emerge with domain edge bands highlighted in red traversing this gap.}
	\label{fig:L=6r=90t05bandsy}
\end{figure}

{\it Band structure:} Since $v_\Delta\ll v_F$, the dispersion along $x$ is almost negligible, Fig. \ref{fig:bandsr90t0x}. For zero twist, there is no gap, Fig. \ref{fig:bandsr90t0y}, but at finite twist, Fig. \ref{fig:L=6r=90t05bandsy}, it opens, with edge modes inside the gap clearly distinguished. While the gap is smaller than for $r=0$, it is comparable in size.

\begin{figure}[h!]
	\centering
	\begin{subfigure}{.45\textwidth}
		\centering
		\includegraphics[width=\linewidth]{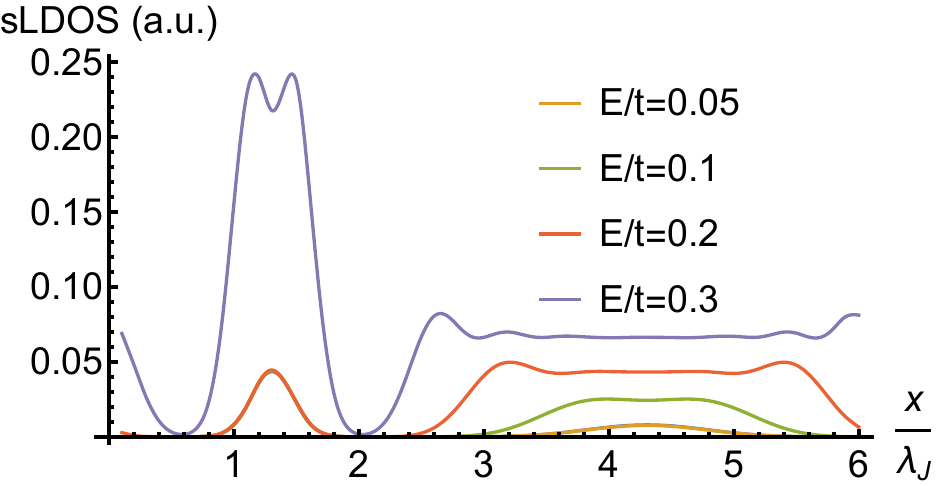}
		\caption{Symmetrized LDOS as a function of real-space coordinate and energy for $\theta=0.5\theta_{MA}$.}
	\end{subfigure}%
	\hspace{3pt}
	\begin{subfigure}{.45\textwidth}
		\centering
		\includegraphics[width=\linewidth]{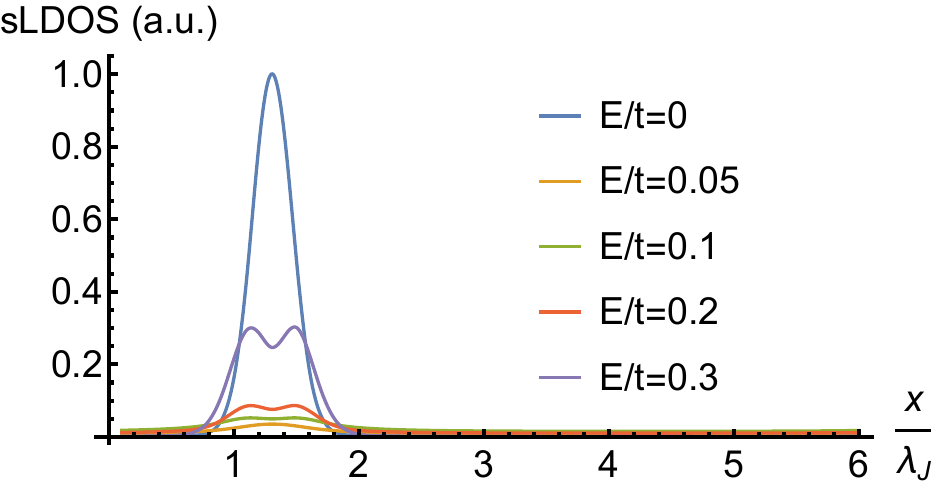}
		\caption{Symmetrized LDOS as a function of real-space coordinate and energy for $\theta=0$.}
	\end{subfigure}
	\caption{Symmetrized LDOS for $r=\pi/2$. LDOS curves almost does not change for a range of energies around $E=0$ for $\theta=0.5\theta_{MA}$, indicating that most of the bands are gapped.}
	\label{fig:L=6r=90ldos}
\end{figure}

{\it LDOS:} In Fig. \ref{fig:L=6r=90ldos} we show energy-symmetrized LDOS. As for $r=0$, for finite twist LDOS does not change at low energies, indicating a gap, in contrast to the $\theta=0$ case. In addition, LDOS at the first peak value $x_{peak}\approx 1.3 \lambda_J$ appears constant for larger range of energies.  This is consistent with the gap to remote bands being larger around $k_y=0$ in Fig. \ref{fig:L=6r=90t05bandsy}.

{\it Calculation details:} "MaxCellMeasure" of $0.03$, a cutoff value $C=3$ for the eigenfunction derivative at $x=l$, level smearing for LDOS calculation$\sigma/t= 0.02$.

\newpage
\subsubsection{Results for $l=6\lambda_J$, $r=\pi/4$}

\begin{figure}[h!]
	\centering
	\begin{subfigure}{.5\textwidth}
		\centering
		\includegraphics[width=\linewidth]{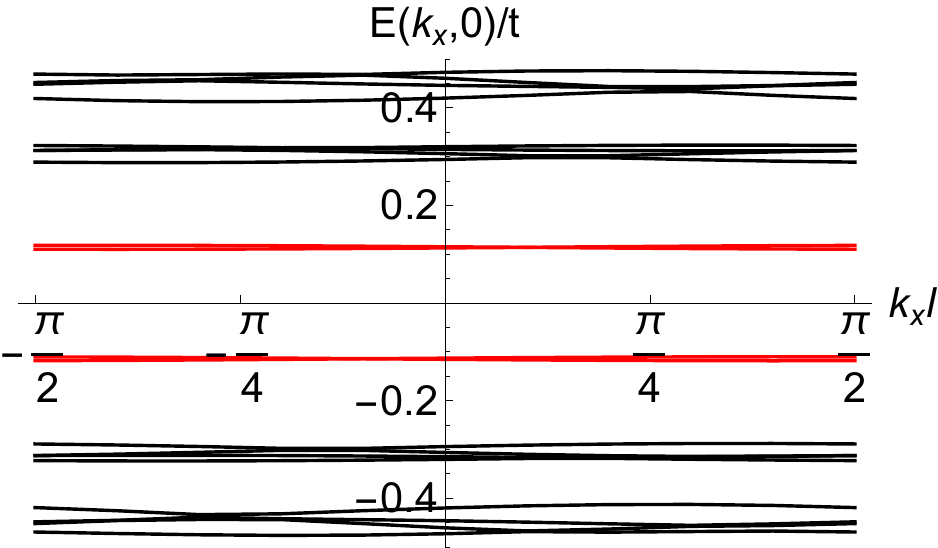}
		\caption{Quasiparticle dispersion along $k_x$.}
		\label{fig:bandsr45t05x}
	\end{subfigure}%
	\begin{subfigure}{.5\textwidth}
		\centering
		\includegraphics[width=\linewidth]{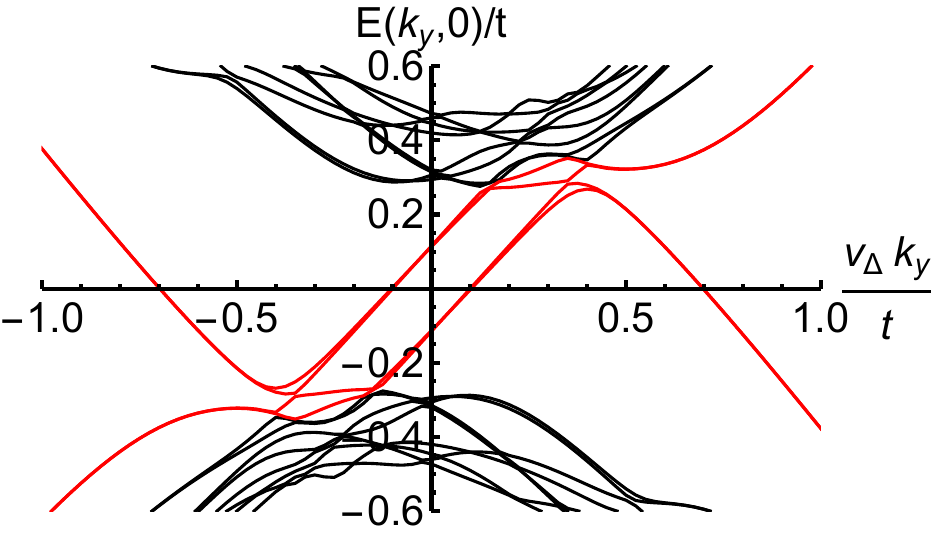}
		\caption{Quasiparticle dispersion along $k_y$.}
		\label{fig:bandsr45t05y}
	\end{subfigure}
	\caption{Quasiparticle band structure in magnetic field for $\theta=0.5 \theta_{MA}$ along (a) $k_x$ and (b) $k_y$ (only few lowest-energy bands are shown).}
	\label{fig:bandsr45t05}
\end{figure}

\begin{figure}[h!]
	\centering
	\begin{subfigure}{.5\textwidth}
		\centering
		\includegraphics[width=\linewidth]{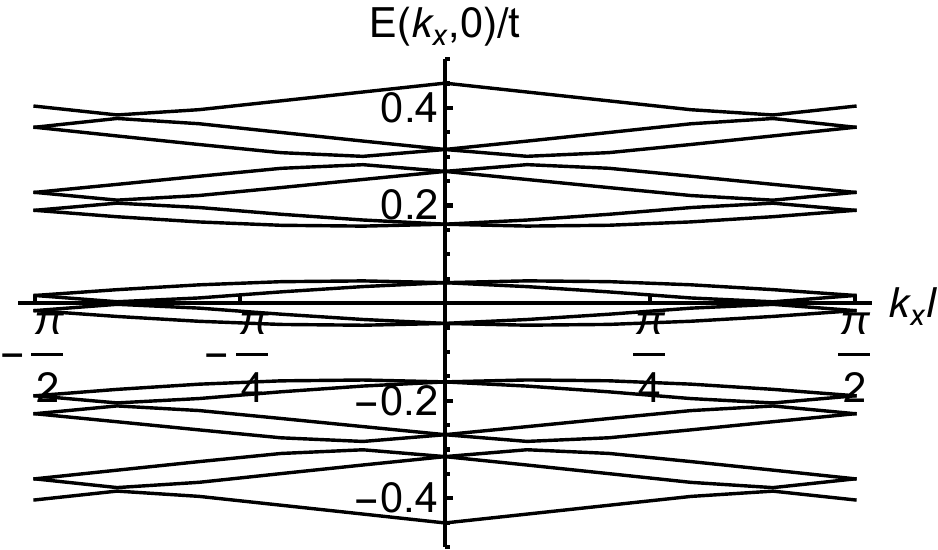}
		\caption{Quasiparticle dispersion along $k_x$.}
		\label{fig:bandsr45t0x}
	\end{subfigure}%
	\begin{subfigure}{.5\textwidth}
		\centering
		\includegraphics[width=\linewidth]{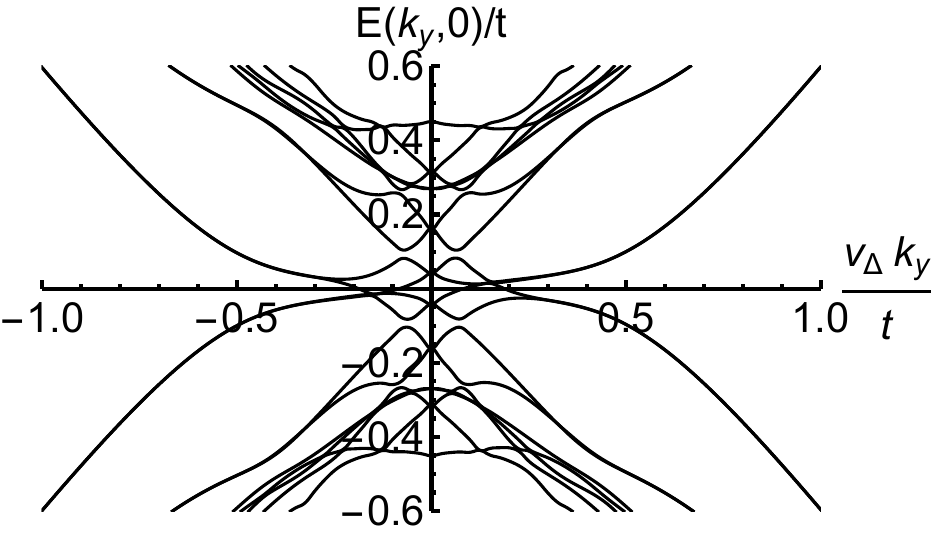}
		\caption{Quasiparticle dispersion along $k_y$.}
		\label{fig:bandsr45t0y}
	\end{subfigure}
	\caption{Quasiparticle band structure in magnetic field for $\theta=0$ along (a) $k_x$ and (b) $k_y$ (only few lowest-energy bands are shown).}
	\label{fig:bandsr45t0}
\end{figure}

{\it Band structure:} The band structure for $\theta=0.5\theta_{MA}$ and $\theta=0$ is shown in Fig. \ref{fig:bandsr45t05} and Fig. \ref{fig:bandsr45t0}, respectively. The results are in good qualitative agreement with $r=0$ case (see main text and above).

\begin{figure}[h!]
	\centering
	\includegraphics[width=0.75\linewidth]{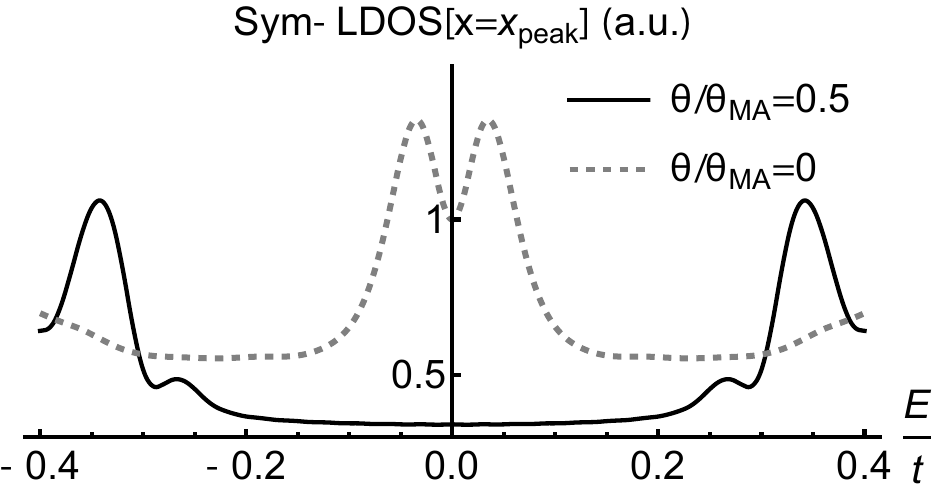}
	\caption{Symmetrized LDOS near $x_{peak}/L\approx 1.305$ for $\theta=0.5\theta_{MA}$. Results are very similar to $r=0$ (see main text).}
	\label{fig:L=6r=45ldospeak}
\end{figure}

{\it LDOS:} In Fig. \ref{fig:L=6r=45ldospeak} we show energy-symmetrized LDOS near $x_{peak}/L\approx 1.305$. As for $r=0$, for finite twist LDOS does not change at low energies, indicating a gap, in contrast to the $\theta=0$ case.

{\it Calculation details:} "MaxCellMeasure" of $0.1$, a cutoff value $C=20$ for the eigenfunction derivative at $x=l$, level smearing for LDOS calculation$\sigma/t= 0.02$. Large $C$ value is to account for stronger dispersion of the bands along $k_y$, a cutoff of $1$ is sufficient at small $k_y$, but at larger $k_y$ even regular eigenvectors would violate such a cutoff.

\newpage

\subsubsection{Results for $l=4 \lambda_J$, $r=0$}
We now discuss how the features change when magnetic field (period of the Josephson vortex lattice) is different. 

\begin{figure}[h!]
	\centering
	\begin{subfigure}{.5\textwidth}
		\centering
		\includegraphics[width=\linewidth]{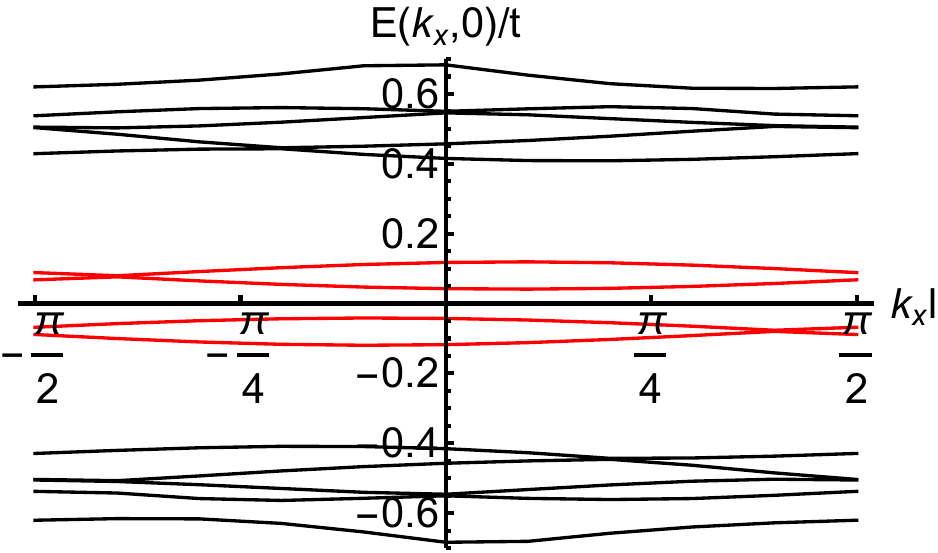}
		\caption{Quasiparticle dispersion along $k_x$.}
		\label{fig:bandsl4x}
	\end{subfigure}%
	\begin{subfigure}{.5\textwidth}
		\centering
		\includegraphics[width=\linewidth]{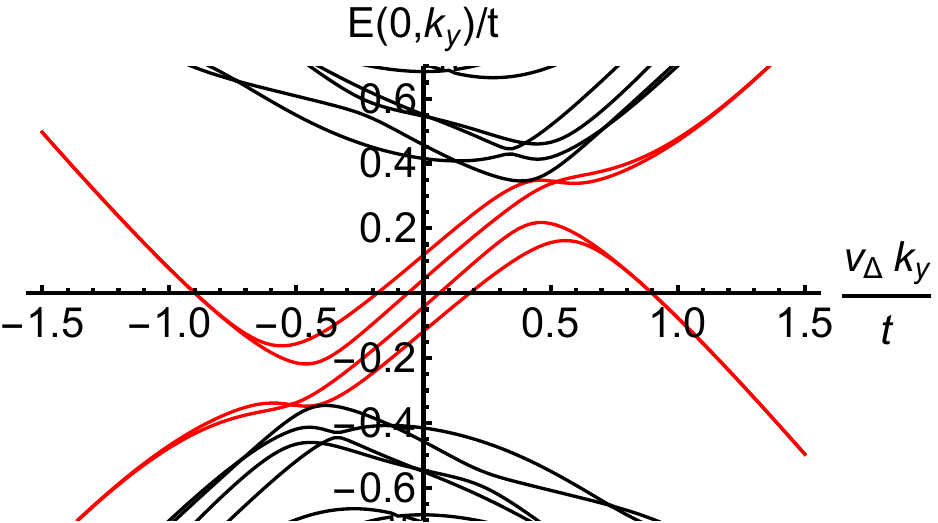}
		\caption{Quasiparticle dispersion along $k_y$.}
		\label{fig:bandsl4y}
	\end{subfigure}
	\caption{Quasiparticle band structure in magnetic field for $\theta=0.5\theta_{MA}$ along (a) $k_x$ and (b) $k_y$ (only few lowest-energy bands are shown).}
	\label{fig:bandsl4}
\end{figure}

\begin{figure}[h!]
	\centering
	\begin{subfigure}{.5\textwidth}
		\centering
		\includegraphics[width=\linewidth]{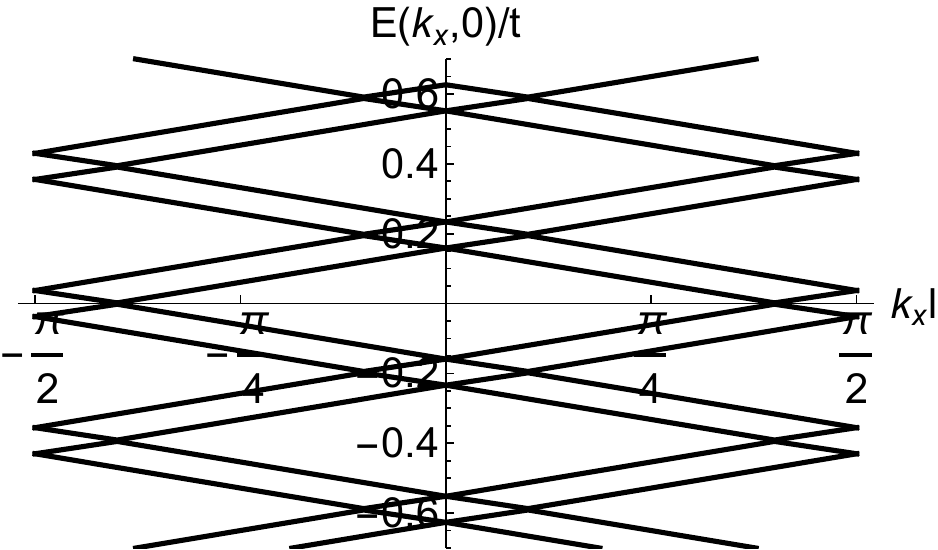}
		\caption{Quasiparticle dispersion along $k_x$.}
		\label{fig:bandsl4xa0}
	\end{subfigure}%
	\begin{subfigure}{.5\textwidth}
		\centering
		\includegraphics[width=\linewidth]{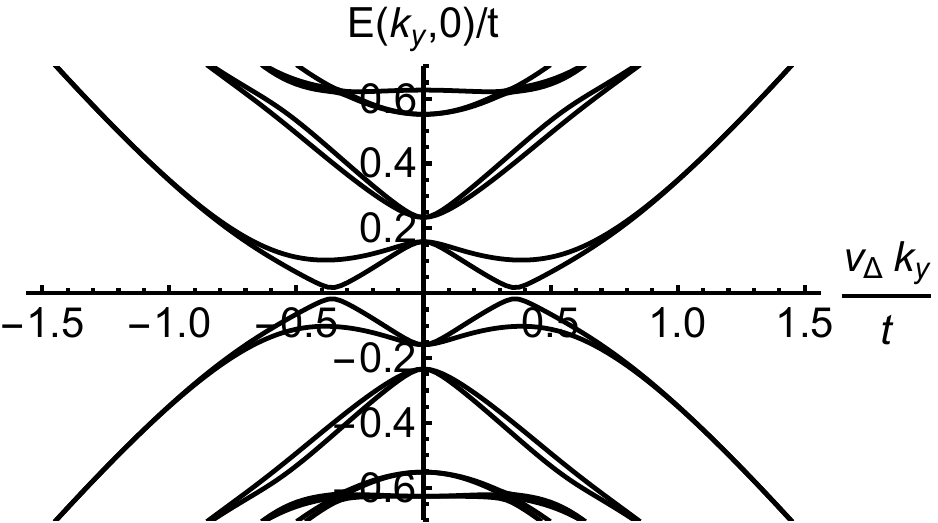}
		\caption{Quasiparticle dispersion along $k_y$.}
		\label{fig:bandsl4ya0}
	\end{subfigure}
	\caption{Quasiparticle band structure in magnetic field for $\theta=0$ along (a) $k_x$ and (b) $k_y$ (only few lowest-energy bands are shown).}
	\label{fig:bandsl4a0}
\end{figure}

{\it Band structure:} The band structure for $\theta=0.5\theta_{MA}$ and $\theta=0$ is shown in Fig. \ref{fig:bandsl4} and Fig. \ref{fig:bandsl4a0}, respectively. The results are in good qualitative agreement with $l=6\lambda_J$ case.

\begin{figure}[h!]
	\centering
	\includegraphics[width=0.75\linewidth]{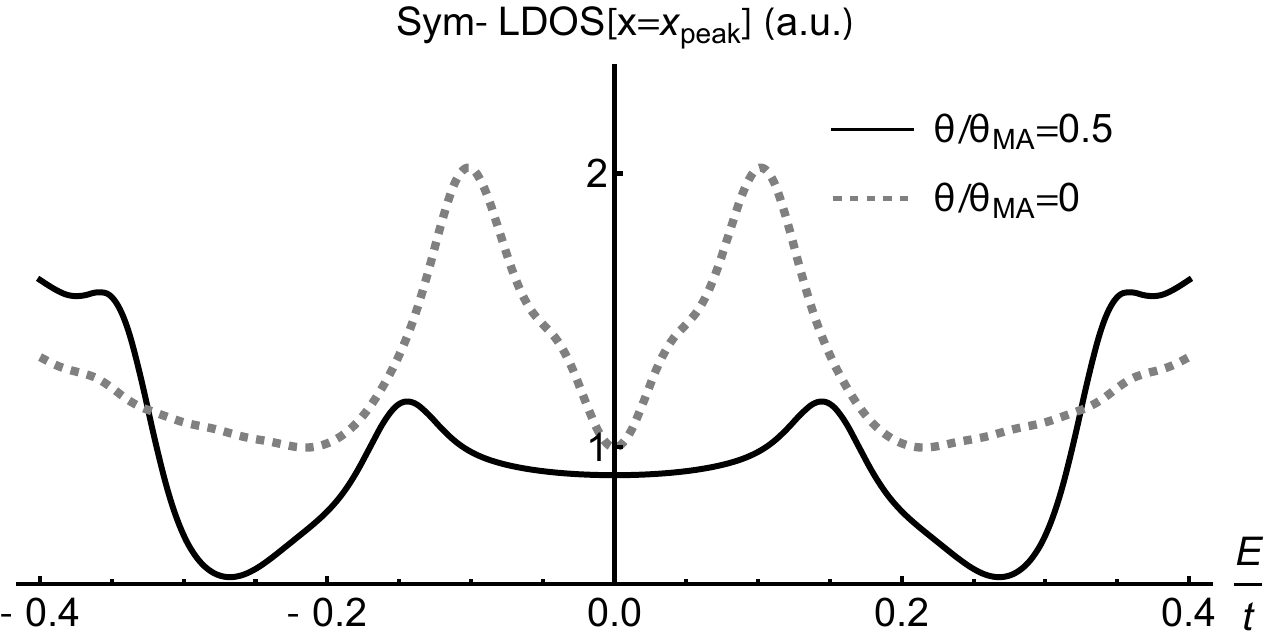}
	\caption{Symmetrized LDOS near $x_{peak}/L\approx 1.305$ for $\theta=0.5\theta_{MA}$. The gap appears smaller than for $l=6\lambda_J$ due to stronger dispersion of bands.}
	\label{fig:L=4r=0ldos}
\end{figure}

{\it LDOS:} In Fig. \ref{fig:L=4r=0ldos} we show energy-symmetrized LDOS near $x_{peak}/L\approx 1.305$. Despite the apparent gap in band structure, the gap in LDOS is somewhat less pronounced due to enhanced dispersion of the bands compared to $l=6\lambda_J$.

{\it Calculation details:} "MaxCellMeasure" of $0.1$, a cutoff value $C=2$ for the eigenfunction derivative at $x=l$, level smearing for LDOS calculation$\sigma/t= 0.025$.

\subsubsection{Results for $l=9 \lambda_J$, $r=0$}
We now discuss how the features change when the period of the Josephson vortex lattice is larger. 

\begin{figure}[h!]
	\centering
	\begin{subfigure}{.5\textwidth}
		\centering
		\includegraphics[width=\linewidth]{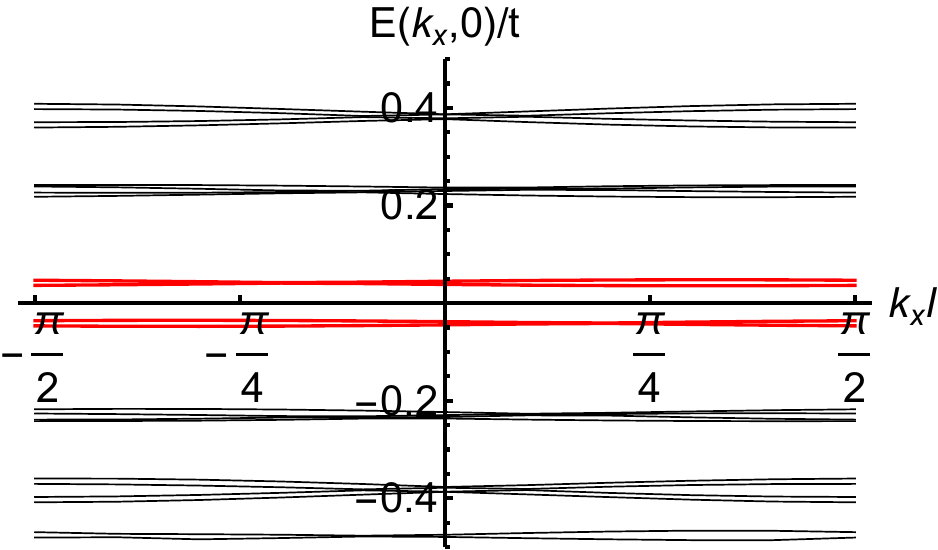}
		\caption{Quasiparticle dispersion along $k_x$.}
		\label{fig:bandsl9x}
	\end{subfigure}%
	\begin{subfigure}{.5\textwidth}
		\centering
		\includegraphics[width=\linewidth]{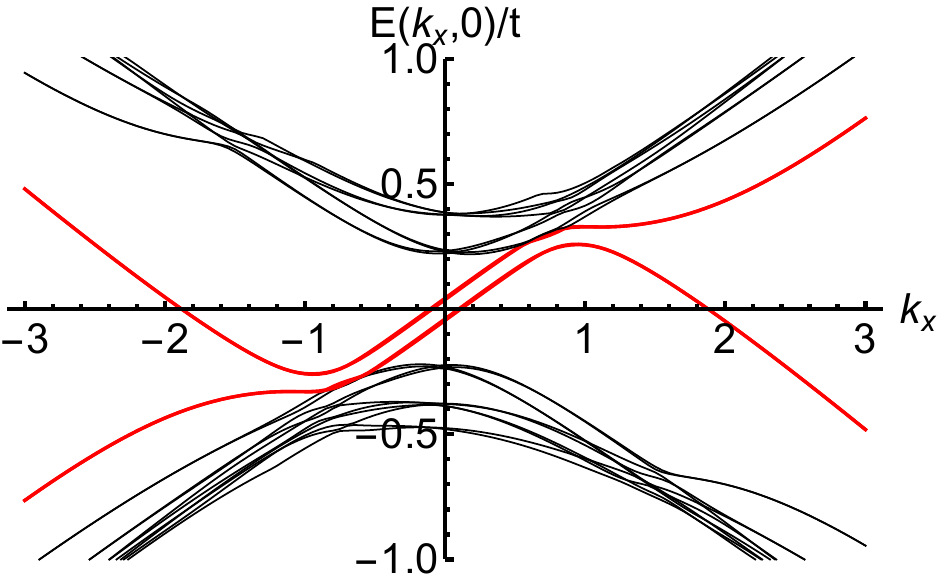}
		\caption{Quasiparticle dispersion along $k_y$.}
		\label{fig:bandsl9y}
	\end{subfigure}
	\caption{Quasiparticle band structure in magnetic field for $\theta=0.5\theta_{MA}$ along (a) $k_x$ and (b) $k_y$ (only few lowest-energy bands are shown).}
	\label{fig:bandsl9}
\end{figure}

{\it Band structure:} The band structure for $\theta=0.5\theta_{MA}$ and $\theta=0$ is shown in Fig. \ref{fig:bandsl9} and Fig. \ref{fig:bandsl4a0}, respectively. The results are in good qualitative agreement with $l=6\lambda_J$ case, but the quasiparticle dispersion along $k_x$ is pronouncedly weaker due to larger separation between vortices.

\begin{figure}[h!]
	\centering
	\begin{subfigure}{.5\textwidth}
		\centering
		\includegraphics[width=\linewidth]{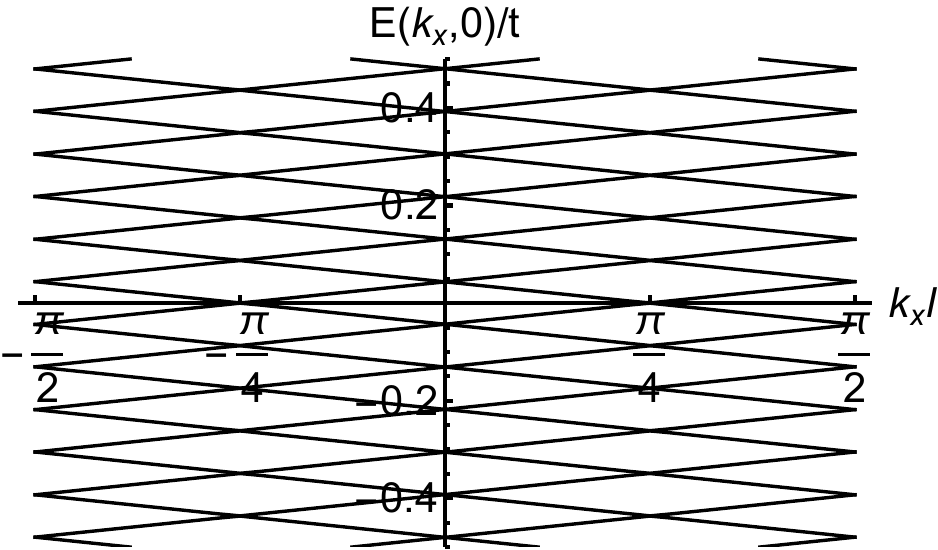}
		\caption{Quasiparticle dispersion along $k_x$.}
		\label{fig:bandsl9xa0}
	\end{subfigure}%
	\begin{subfigure}{.5\textwidth}
		\centering
		\includegraphics[width=\linewidth]{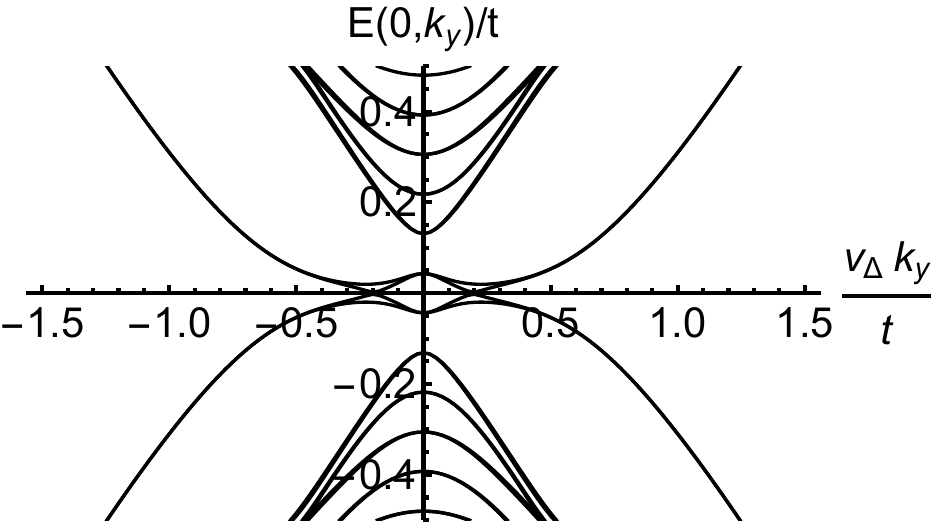}
		\caption{Quasiparticle dispersion along $k_y$.}
		\label{fig:bandsl9ya0}
	\end{subfigure}
	\caption{Quasiparticle band structure in magnetic field for $\theta=0$ along (a) $k_x$ and (b) $k_y$ (only few lowest-energy bands are shown).}
	\label{fig:bandsl9a0}
\end{figure}
\begin{figure}[h!]
	\centering
	\includegraphics[width=0.75\linewidth]{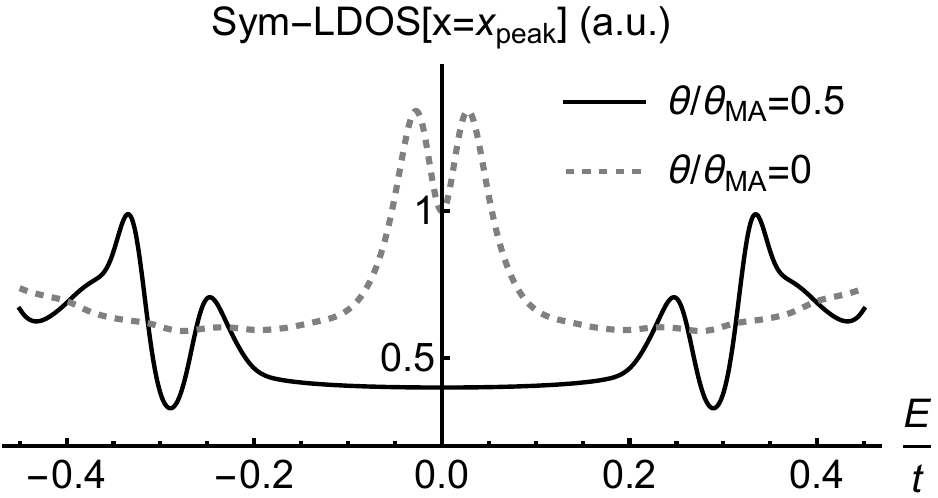}
	\caption{Symmetrized LDOS near $x_{peak}/L\approx 1.718$ for $\theta=0.5\theta_{MA}$.}
	\label{fig:L=9r=0ldos}
\end{figure}

{\it LDOS:} In Fig. \ref{fig:L=9r=0ldos} we show energy-symmetrized LDOS near $x_{peak}/L\approx1.718$. The result compares well to the case $l=6\lambda_J$.

{\it Calculation details:} "MaxCellMeasure" of $0.2$, a cutoff value $C=1$ for the eigenfunction derivative at $x=l$, level smearing for LDOS calculation $\sigma/t= 0.02$.

\newpage
\subsection{Analytical approximation for $\alpha\ll1$}
To describe the localized domain-edge modes we develop an analytical approximation near $\varphi \mod 2\pi \approx 0,\pi$ and $\alpha\ll1$. For $\varphi\approx 0$, one can use the Dirac point approximation of the main text, with details given in\footnotemark[\value{footnote}]. One additional term arising due to the magnetic field, $h\sigma_3 \tau_3$ projects to $\pm \alpha h \zeta_1$ in the Dirac point basis. Furthermore, we expand $\frac{\sin \varphi(x)}{2} \approx \frac{q_0 x}{2}$, where $q_0\equiv \varphi'(x_0)$, where $\varphi(x_0,x_\pi) \mod 2 \pi = 0,\pi$. The Schrodinger equation takes then the form:
\begin{equation}
	\left[v_F (-i\partial_x) \zeta_3 + (v_\Delta k_y\pm \alpha h) \zeta_1 + \frac{\alpha q_0 x t}{2} \zeta_2\right] \psi_n(x) = E_n \psi_n(x),
	\label{eq:hamphi=0}
\end{equation}
where $ \psi_n(x) = [\psi_n^{(1)}(x),\psi_n^{(2)}(x)]$. To simplify notation we shift $k_y=k_y'\mp \alpha h/ v_\Delta$. Applying $[\zeta_3 v_F (-i\partial_x)-E_n]$ to the resulting equation one gets:
\begin{equation}
	[(v_F^2 \partial_{xx}+ E_n^2- (v_\Delta k_y')^2-(\alpha t q_0 x)^2/4)+\alpha t q_0/2 \zeta_1 ]\psi_n(x) = 0,
\end{equation}
which has solution in terms of oscillator eigenstates:
\begin{equation}
	\begin{gathered}
		\psi_n^{\pm}(x) = [1/\sqrt{2},\pm 1/\sqrt{2}] \psi^{osc}_n(x),
		\\
		\psi^{osc}_n(x) = \frac{1}{\sqrt{2^n n!}} \left(\frac{\alpha t q_0}{2 \pi v_F}\right)^{1/4}e^{-\frac{\alpha t q_0 x^2}{4 v_F}} H_n\left( \sqrt{\frac{\alpha t q_0}{2 v_F}} x \right)
	\end{gathered}
\end{equation}
which yields eigenenergy equation $(E_n^\pm)^2= v_F q_0 \alpha t (n+1/2) \mp \alpha t q_0/2 + (v_\Delta k_y')^2$. Except $(E_0^{+})^2= (v_\Delta k_y')^2$, all other energies are doubly degenerate. Therefore, eigenfunctions take the form $a\psi_{n+1}^{+}(x) + b \psi_n^{-}(x)$ for $n\geq 0$. From Eq. \eqref{eq:hamphi=0} one finds the resulting eigenvalues and eigenvectors:
\begin{equation}
	\begin{gathered}
		E_0 = v_\Delta k_y \pm \alpha h;\; \psi_0(x) = [1/\sqrt{2}, 1/\sqrt{2}] \psi^{osc}_0(x);
		\\
		E_n^{\pm} = \pm\sqrt{v_F q_0 \alpha t n + (v_\Delta k_y \pm  \alpha h)^2};\;
		\\
		\psi_n^{\pm}(x) = a [1/\sqrt{2}, 1/\sqrt{2}] \psi^{osc}_{n+1}(x)+ b[1/\sqrt{2},- 1/\sqrt{2}] \psi^{osc}_{n}(x);
		\\
		\frac{b}{a} = \frac{-i \sqrt{v_F \alpha t q_0(n+1)}}{E_n^{\pm}+v_\Delta k_y\pm  \alpha h}.
	\end{gathered}
	\label{eq:eigphi=0}
\end{equation}

\begin{figure}[h!]
	\centering
	\includegraphics[width=0.75\linewidth]{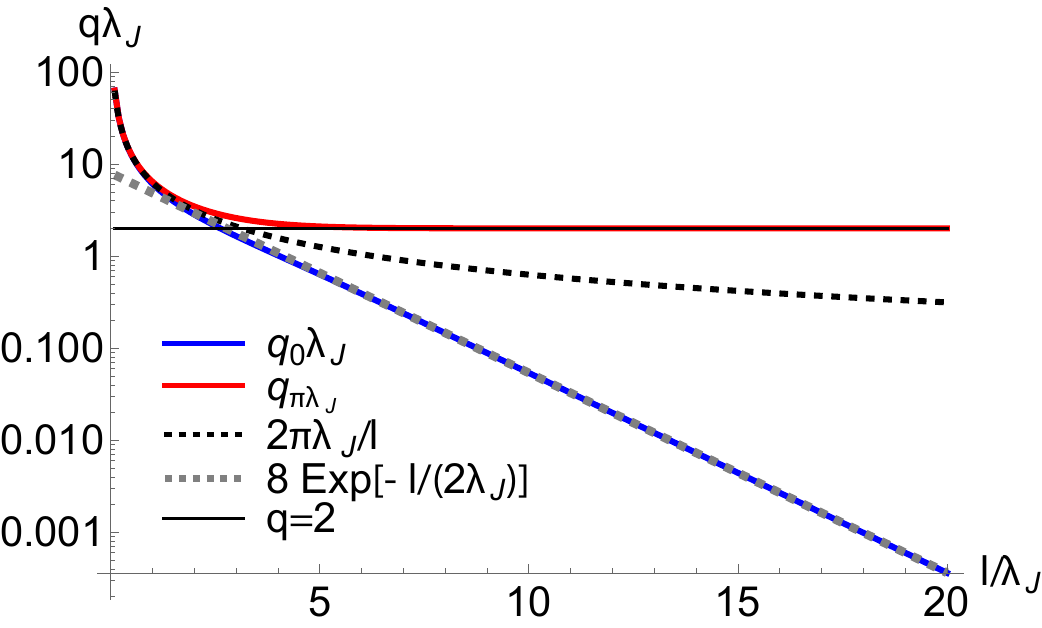}
	\caption{$q_{0,\pi} = \varphi'(x_0,x_\pi)$ as a function of $l$, where $\varphi(x_0,x_\pi) \mod 2 \pi = 0,\pi$. The exponential behavior of $q_0$ at $l\gg\lambda_J$ indicates the increasing separation between highly nonlinear Josephson vortices, such as the current between them is exponentially suppressed. $q_\pi$, on the other hand, saturates at $2/\lambda_J$, because $x_\pi$ is at the center of the Josephson vortex which keeps its shape as the distance between vortices increases. Finally, at $l\ll\lambda_J$ $\varphi(x)\approx 2\pi x/l$, as the l.h.s. of Eq. \eqref{eq:josephson} dominates in this regime.}
	\label{fig:qvsl}
\end{figure}

Most importantly, Eq. \eqref{eq:eigphi=0} allows one to find the gap between the chiral domain mode and the intra-domain states: it is equal to $\sqrt{v_F q_0 \alpha t}$. Let us now discuss the applicability of the expansion $\sin (\varphi(x))\approx q_0(x-x_0)$. The characteristic wavefunction size is given by $\sqrt{\langle x^2 \rangle}= \sqrt{v_F/(\alpha t q_0)}$. Therefore, the approximation is valid for $\varphi'(x_0) \sqrt{\langle x^2 \rangle}  = \sqrt{\frac{v_F q_0}{\alpha t}} \ll 1$. In Fig. \ref{fig:qvsl}  we plot the value of $q_0$ as a function of $l$ - it is clearly decreasing exponentially at $l\gg\lambda_J$ reflecting the increasing separation between highly nonlinear Josephson vortices. Therefore, the results presented here hold for $l\gg \lambda_J$.

For  $x \approx x_\pi$ we first need to analyze the low-energy excitations for $\varphi \approx \pi$. For $\varphi=\pi$ the Hamiltonian, Eq. 1 of the main text, takes the form:

\begin{equation}
	\begin{gathered}
		H({\bf k},\varphi \approx \pi) = H^{0}({\bf k})+H^{1}({\bf k})(\varphi-\pi)+O[(\varphi-\pi)^2];
		\\
		H^{0}({\bf k})
		=	
		v_F k_\parallel\tau_3 \sigma_1+ t\tau_3\sigma_1 - v_\Delta k_\perp  \tau_2 \sigma_3 +\alpha t;
		\\
		H^{1}({\bf k})
		=
		- \frac{v_\Delta k_\perp}{2} \tau_1+ \frac{\alpha t}{2} \sigma_3 \tau_1.
	\end{gathered}
	\label{eq:hamKphipi}
\end{equation}
$H^{0}({\bf k})$ has zero eigenvalues at $k_\parallel = 0; v_\Delta k_\perp = \pm t \sqrt{\alpha^2+1}$ with eigenvectors:
\begin{equation}
	\begin{gathered}
		e_1^\pm = [i (\alpha\pm \sqrt{\alpha^2+1}),0,0,1]/\sqrt{1+(\alpha\pm \sqrt{\alpha^2+1})^2};\;
		\\
		e_2^\pm = [0,i (\alpha\pm \sqrt{\alpha^2+1}),1,0]/\sqrt{1+(\alpha\pm \sqrt{\alpha^2+1})^2}.
	\end{gathered}
	\label{eq:eigenvectorsphi=pi}
\end{equation}
Projecting $H_0,H_1$, Eq. \ref{eq:hamKphipi} to the basis of Eq. \ref{eq:eigenvectorsphi=pi} one gets:
\begin{equation}
	\begin{gathered}
		H({\bf k},\varphi \approx \pi) 
		\approx 
		\pm \frac{\alpha}{\sqrt{\alpha^2+1}} v_F k_\parallel \zeta_3 - v_\Delta k_y \zeta_2	\mp \frac{t(\varphi-\pi)}{2\sqrt{\alpha^2+1}} \zeta_1.
	\end{gathered}
	\label{eq:hamKphipiappr}
\end{equation}
This Hamiltonian is gapped for $\varphi \neq \pi$. Importantly, the Chern number for $\varphi<\pi$ matches the one in the main text $\varphi\approx 0$, while for $\varphi>\pi$ it is opposite, indicating a topological transition with gap closing at $\varphi=\pi$ and then reopening.

We can now take into account the varying phase and magnetic field. The latter, equal to a term $-h\tau_3\sigma_3$ projects to $-h\zeta_3$ for both nodes. Close to $x=x_\pi,k_y = \pm t\sqrt{\alpha^2+1}$ and expanding for $\alpha\ll 1$ one gets:
\begin{equation}
	H_{\varphi(x)\approx \pi} 
	=
	\pm (\alpha v_F (-i\partial_x) - h) \zeta_3 - v_\Delta k_y \zeta_2	\mp \frac{q_\pi x t}{2} \zeta_1,
	\label{eq:hamphipi}
\end{equation}
where $q_\pi=\varphi'(x_\pi)$. One observes that magnetic field effect can be absorbed into $\psi(x) \to e^{i h x/(\alpha v_F)}\psi(x)$. Furthermore, a transformation $U=\zeta_2$ makes the Hamiltonian near two nodes identical and an additional rotation $U' = \frac{1+i\zeta_3}{\sqrt{2}}$ bring the Hamiltonian to the form:

\begin{equation}
	\alpha v_F (-i\partial_x)  \zeta_3 - v_\Delta k_y \zeta_1	+\frac{q_\pi x t}{2} t \zeta_2
\end{equation}
that is identical in form to \ref{eq:hamphi=0} with $k_y\to -k_y$. Therefore, the eigenstates and eigenvalues of \eqref{eq:hamphipi} can be obtained from \eqref{eq:eigphi=pi} by substituting $v_F\to \alpha v_F;\; k_y\to-k_y\; q \to q_\pi/\alpha$ and applying $\frac{1-i\zeta_3}{\sqrt{2}}, k_y\to-k_y$ to the eigenfunctions. As a result the eigenvalues are:
\begin{equation}
	\begin{gathered}
		E_0 = - v_\Delta k_y ;\\
		E_n^{\pm} = \pm\sqrt{\alpha v_F q_\pi t n + (v_\Delta k_y )^2};
	\end{gathered}
	\label{eq:eigphi=pi}
\end{equation}
while the zeroth level eigenfunction, describing the chiral domain edge mode band is given by:
\begin{equation}
	\tilde{\psi}_0(x) = [(1-i)/2, (1+i)/2] \left(\frac{t q_\pi}{2 \pi \alpha v_F}\right)^{1/4}e^{-\frac{t q_\pi x^2}{4 \alpha v_F}}.
\end{equation}
The characteristic wavefunction size is given by $\sqrt{\langle x^2 \rangle}= \sqrt{\alpha v_F/(t q_\pi)}$, such that the approximate Hamiltomnian is valid for $\sqrt{\alpha v_F q_\pi/t}\ll1$.

For $l/\lambda_J=6$ used in the main text, the wavefunction sizes are $0.35 \lambda_J$ for $x\approx x_\pi$ and $1.60\lambda_J$ for $x\approx x_0$, compared to $0.54\lambda_J$ and $1.25\lambda_J$ in numerical calculations. Given that dimensionless parameters for the analytical approximation are $0.7$ and $0.6$, respectively, the agreement is reasonable.

\section{Disorder}

Here we consider the robustness of the current-induced topological gap in the presence of disorder. Disorder potential is described by:

\begin{equation}
	H_{imp} = \int d {\bf r} V_{imp}({\bf r})  c^\dagger({\bf r})\hat{T}c({\bf r}),
	\quad
	V_{imp}({\bf r}) =  u_T\sum_i \delta({\bf r}-{\bf r}_i),
\end{equation}
where $\hat{T}$ is a hermitian matrix describing the structure of impurity potential in Gor'kov-Nambu space (e.g. $\tau_3$ for ordinary charge impurity or $\tau_1$ for local gap suppression \cite{hettler1999,pereg2008}) and layer space (e.g. $(1\pm\sigma_3)/2$ for impurities in a single layer); an average over the impurity positions ${\bf r}_i$ is to be taken in the end of the calculation. We restrict our analysis to Born approximation, assuming weak impurity potential $u_0 k_N^3\ll E_F,\Delta_0$ \cite{agd}. While rare region effects or multiple scattering, ignored in the Born approximation, can lead to creation of bound states within the topological gap, these will be localized and not affect the transport properties for impurity concentration below a critical one.

We start with analyzing the effect of disorder in the low-energy Dirac node approximation discussed in the main text:

\begin{equation}
	H_{\mathrm{eff}}({\bf k},\varphi)
	=\tilde {v}_{F}(k_\parallel-k_\parallel^N)\zeta_3
	+ \alpha t \sin(\varphi/2)\zeta_2 
	+\tilde {v}_{\Delta}\cos(\varphi/2)k_\perp\zeta_1.
\end{equation}

Note that there are two Dirac nodes per valley, with different low-energy basis \footnotemark[\value{footnote}]:  $\{(|e_1\rangle+|e_2\rangle)/\sqrt{2},(|e_1\rangle-|e_2\rangle)/\sqrt{2}\}$ around $\xi^N= \sqrt{1-\alpha^2} t,\delta^N=0$ and $\{(|e_1\rangle+|e_2\rangle)/\sqrt{2},(-|e_1\rangle+|e_2\rangle)/\sqrt{2}\}$ around $\xi^N= -\sqrt{1-\alpha^2} t,\delta^N=0$ for $|\alpha|<1$, where:
\begin{equation}
	\begin{gathered}
		|e_1\rangle =[-\xi^N,\delta_0-t,t-\delta_0,\xi^N]^T
		/(2t\sqrt{t-\delta_0})
		,
		\\
		|e_2\rangle =[-\xi^N,t+\delta_0,t+\delta_0,-\xi^N]^T
		/(2t\sqrt{t+\delta_0})
		,
	\end{gathered}
\end{equation}
where $\delta_0 \equiv\frac{{\bf v}_\Delta \cdot {\bf Q}_N}{2}$. Eigenvectors for adjacent valleys, as described in text, follows from $\delta_0\to -\delta_0$. The projections of relevant Pauli matrices in $\hat{T}$ to the Dirac point basis are given in Table \ref{tab}.

\begin{table}
	\centering
	\begin{tabular}{cccccc}
		\hline \hline
		Type of scattering & $\tau_3$ & $\sigma_3\tau_3$& $\sigma_1\tau_3$&   $\tau_1$ &   $\sigma_3\tau_1$\\
		\hline 
		Intranode intravalley & $\sqrt{1-\alpha^2}\zeta_3$ & $\pm\alpha \zeta_1$ & $\mp\zeta_3$& $\sqrt{1-\alpha^2} \zeta_1$& $\mp\alpha \zeta_3$\\
		Internode intravalley & $0$ & $-\zeta_3$ & $-\alpha \zeta_1$ & 0 & $-\zeta_1$ \\
		Intranode intervalley &$\zeta_3$ &$0$ & $\mp\sqrt{1-\alpha^2}\zeta_3$& $\zeta_1$ & $0$ \\
		Internode intervalley & $\mp \alpha \zeta_3$ &$-\zeta_3$ & 0  &$\pm\alpha\zeta_3$&$-\sqrt{1-\alpha^2}\zeta_1$\\
	\end{tabular}
	\caption{Projection of various Pauli matrices, relevant for impurity potential, to the basis of}
	\label{tab}
\end{table}

Let us consider first the lowest-order self energy due to disorder:
\begin{equation}
	\Sigma_0(i\varepsilon_n)  = - n u_0^2 \int \frac{d k_\parallel d k_\perp}{(2\pi)^2}  \hat{T}_{proj}  \frac{i\varepsilon_n+v_F k_\parallel \zeta_3+v_\Delta k_\perp \zeta_1+\Delta_J \zeta_2}{\varepsilon_n^2+(v_F k_\parallel)^2+(v_\Delta k_\perp)^2+\Delta_J^2} \hat{T}_{proj},
	\label{eq:sigmaborn0}
\end{equation}
where $n$ is the impurity concentration. Two middle terms in Eq. \eqref{eq:sigmaborn0} vanish after $k$ integration, and for charge ($\hat{T}\sim \tau_3$)/gap($\hat{T}\sim \tau_1$) impurities that do not break layer symmetry on average ($\hat{T}_{proj}^2\sim \sigma_0$) the self-energy takes the general form $\Sigma_0(i\varepsilon_n) = \Sigma_0^0(i\varepsilon_n)+\Sigma_0^2(i\varepsilon_n) \tau_2$. Therefore, we can evaluate the sum of all non-crossing diagrams (while the crossing ones are small in the Born limit \cite{agd}) as:
\begin{equation}
	\begin{gathered}
		\Sigma(i\varepsilon_n) = i(\varepsilon_n -\tilde{\varepsilon}_n) + (\tilde{\Delta}_J(\varepsilon_n)-\Delta_J)\zeta_2  =
		\\
		=
		- n u_0^2 \int \frac{d k_\parallel d k_\perp}{(2\pi)^2}  \hat{T}_{proj}  \frac{i\tilde{\varepsilon}_n+v_F k_\parallel \zeta_3+v_\Delta k_\perp \zeta_1+\tilde{\Delta_J} \zeta_2}{\tilde{\varepsilon}_n^2+(v_F k_\parallel)^2+(v_\Delta k_\perp)^2+\tilde{\Delta_J}^2} \hat{T}_{proj},
		\label{eq:sigmaborn}
	\end{gathered}
\end{equation}
with explicit equations:
\begin{equation}
	\begin{gathered}
		\varepsilon_n = \tilde{\varepsilon}_n \left(1-n u_0^2 \int \frac{d k_\parallel d k_\perp}{(2\pi)^2}  \frac{{\rm Tr} [\hat{T}^2_{proj}]/2}{\tilde{\varepsilon}_n^2+(v_F k_\parallel)^2+(v_\Delta k_\perp)^2+\tilde{\Delta_J}^2} \right),\\
		\Delta_J =\tilde{\Delta_J} \left(1+n u_0^2 \int \frac{d k_\parallel d k_\perp}{(2\pi)^2}  \frac{{\rm Tr} [\zeta_2 \hat{T}_{proj}\zeta_2 \hat{T}_{proj}]/2}{\tilde{\varepsilon}_n^2+(v_F k_\parallel)^2+(v_\Delta k_\perp)^2+\tilde{\Delta_J}^2} \right),
		\label{eq:sigmaborneqn}
	\end{gathered}
\end{equation}
that can be evaluated to yield:
\begin{equation}
	\begin{gathered}
		\varepsilon_n = \tilde{\varepsilon}_n \left(1- g_0\log\frac{\Lambda^2}{\tilde{\varepsilon}_n^2+\tilde{\Delta_J}^2}\right),\\
		\Delta_J =\tilde{\Delta_J}  \left(1- g_1\log\frac{\Lambda^2}{\tilde{\varepsilon}_n^2+\tilde{\Delta_J}^2}\right),\\
		g_0 =  \frac{n u_0^2 {\rm Tr} [\hat{T}^2_{proj}]/2 }{4\pi v_F v_\Delta}
		,\;
		g_1 = - \frac{n u_0^2 {\rm Tr} [\zeta_2 \hat{T}_{proj}\zeta_2 \hat{T}_{proj}]/2 }{4\pi v_F v_\Delta}
		\label{eq:sigmaborneqn2}
	\end{gathered}
\end{equation}
with $\Lambda$ being the upper cutoff in energy (of the order $\Delta_0$). Note that while for a single valley with $\hat{T}_{proj} \sim \zeta_{1,2}$, $g_1=g_0$ this is not true if scattering to other valleys is included. Indeed, as discussed in the main text, for adjacent valleys $\Delta_J$ have opposite signs. Therefore for internode scattering $g_1\to-g_1$ while $g_0\to g_0$ and one expects $g_1<g_0$ after summation over all valleys.

The quantity of interest is the density of states, which is zero within the gap without disorder. Without loss of generality we will assume ${\rm Tr} [\hat{T}^2_{proj}]/2=1$. Density of states is then given by \cite{mineev1999introduction} (assuming spin rotation symmetry is preserved):
\begin{equation}
	\begin{gathered}
		N(\omega) = - \frac{1}{\pi} \int \frac{d k_\parallel d k_\perp}{(2\pi)^2} {\rm Im} {\rm Tr} \left[ \frac{1+\sqrt{1-\alpha^2} \zeta_3}{2} G(i\varepsilon_n, {\bf k})_{i\varepsilon_n\to \omega+i0}\right]
		=
		\\
		=
		-\frac{1}{n u_0^2 \pi} {\rm Im} 
		{\rm Tr} \left[ \frac{1+\sqrt{1-\alpha^2} \zeta_3}{2} \Sigma(i\varepsilon_n)_{i\varepsilon_n\to \omega+i0}\right]
		=
		\frac{1}{n u_0^2 \pi} {\rm Im} 
		(i\tilde{\varepsilon}_n)_{i\varepsilon_n\to \omega+i0},
	\end{gathered}
	\label{eq:dosdisord}
\end{equation}
where we used $\tau_{3,proj} = \sqrt{1-\alpha^2} \zeta_3$.

Let us first consider the case of zero interlayer current $\Delta_J=\tilde{\Delta_J} = 0$. The corrections due to disorder in \eqref{eq:sigmaborneqn2} are singular at $\varepsilon_n=0$ and cannot be taken as small. Instead, one can obtain an approximate solution for small $\varepsilon_n$, noticing that the term inside the bracket can vanish:
\begin{equation}
	\tilde{\varepsilon}_n|_{\Delta_J=0} = 	\Lambda e^{-1/(2g_0)} + \frac{\varepsilon_n}{2g_0},
\end{equation}
which leads to a finite DOS at zero energy $N(\omega=0) =\frac{\Lambda e^{-1/(2g_0)}}{n u_0^2 \pi}$, while Dirac points in a clean system would yield zero. Thus, Dirac points are not stable with respect to infinitesimal disorder.

In contrast to that, for finite $\Delta_J$ infinitesimal disorder does not lead to a finite DOS at zero energy. Indeed, for $g_{0,1}\ll \log^{-1} \frac{\Lambda^2}{\Delta_J^2}$ one has $\tilde{\varepsilon}_n\approx \varepsilon_n \left(1- g_1\log\frac{\Lambda^2}{\Delta_J^2}\right)^{-1}$ for $\varepsilon_n\ll\Delta_J$ and thus $N(0)=0$. Finite density of states appears first above a critical disorder strength $g_0^{cr}$, where the solution for small $\epsilon_n$ is given by:
\begin{equation}
	\begin{gathered}
		\tilde{\Delta}_J \approx \frac{\Delta_J}{1-g_1/g_0},
		\\
		\tilde{\varepsilon}_n \approx \sqrt{\Lambda^2 e^{-1/g_0}-\tilde{\Delta}^2} + \varepsilon_n \frac{\Lambda^2 e^{-1/g_0}}{2g_0(\Lambda^2 e^{-1/g_0}-\tilde{\Delta}^2)},
		\\
		g_0 > g_0^{cr}> 1/\log \left[\frac{(1-g_1/g_0)^2\Lambda^2}{\Delta_J^2}\right],
		\\
		N(0) = \frac{ \sqrt{\Lambda^2 e^{-1/g_0}-\tilde{\Delta}^2} }{n u_0^2 \pi}.
	\end{gathered}
\end{equation}
Therefore, the current-induced spectral gap is robust to weak disorder.

\begin{figure}[h!]
	\centering
	\includegraphics[width=0.75\linewidth]{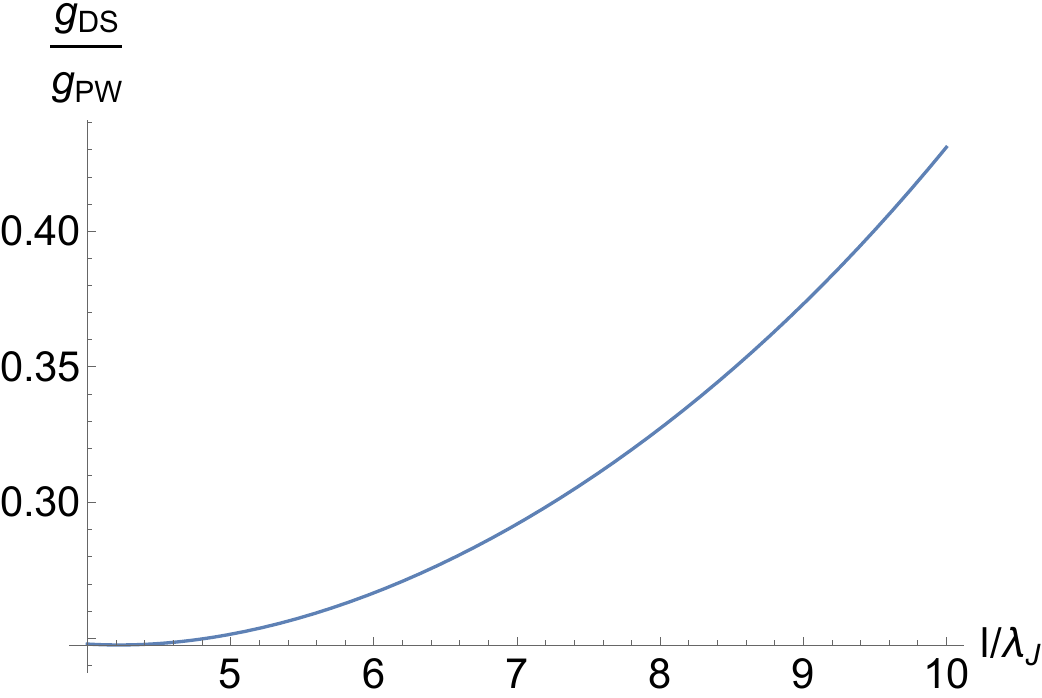}
	\caption{Scattering rate of adjacent domain wall state due to disorder, relative to that of plane wave states for $\alpha=0.5, \tilde{v}_F = 0.5$. At all values of $l$, $\sqrt{v_F/(\alpha t q_0)}/l<0.5$ and $\sqrt{v_F q_0/(\alpha t)}<1$, such that the localized approximation \eqref{eq:eigphi=0} can be used at least qualitatively.}
	\label{fig:scrate}
\end{figure}

Finally, let us comment on the effects of disorder on the field-induced chiral domain modes. Indeed, scattering between modes with opposite chirality along $y$ may lead to a gap opening. However, the modes are localized in $x$ direction at different positions, and the resulting potential for scattering will be strongly reduced by the overlap of domain wall state wavefunctions. In particular, for point-like impurities the scattering rate is proportional to the absolute value of the product of eigenfunctions squared at the impurity position. Normalizing wavefunctions within one vortex lattice unit cell one gets the ratio of scattering amplitudes for adjacent domain states (DS) and plane waves (PW):
\begin{equation}
	\frac{g_{DS}}{g_{PW}} = \left\langle\left\langle \frac{ |\psi^*_{x\approx x_0}(x_{imp}) \psi'_{x\approx x_\pi}(x_{imp})|^2}{1/l^2} \right\rangle\right\rangle_{x_{imp}},
\end{equation}
where $\langle\langle ... \rangle\rangle_{x_{imp}}$ represents the average over impurity positions along $x$. Taking into account $\alpha q_0\ll q_\pi/\alpha$ one gets 
\begin{equation}
	\langle\langle |\psi^*_{x\approx x_0}(x_{imp}) \psi'_{x\approx x_\pi}(x_{imp})|^2 \rangle\rangle_{x_{imp}}
	\approx
	|\psi^*_{x\approx x_0}(x_{\pi})|^2   \int \frac{dx_{imp}}{l} |\psi'_{x\approx x_\pi}(x_{imp})|^2 =
	|\psi^*_{x\approx x_0}(x_{\pi})|^2 /l.
\end{equation}
Thus, one finally obtains:
\begin{equation}
	\frac{g_{DS}}{g_{PW}} \approx  (l/\lambda_J) \left(\frac{\alpha (q_0(l)\lambda_J) }{2 \pi \tilde{v}_F}\right)^{1/2}e^{-\frac{\alpha (q_0(l)\lambda_J) (l/2\lambda_J)^2}{2 \tilde{v}_F}}.
\end{equation}
In Fig. \ref{fig:scrate} $\frac{g_{DS}}{g_{PW}}$ is presented as a function of $l$ for $\alpha=0.5, \tilde{v}_F = 0.5$. One observes that the scattering rate reduction can be substantial, e.g. $0.27$ for $l=6\lambda_J$ used in the main text.

\end{document}